\documentclass[]{emulateapj}
\usepackage{graphics}
\usepackage{color,ulem,epstopdf}

\newcommand{\y}{}
\newcommand{\bc}{}

\newcommand\icarus{\ref@jnl{Icarus}}

\shorttitle{Planetary Evaporation}
\shortauthors{Owen, J.E. \& Wu, Y.}

\begin{document}


\title{{\it Kepler} planets: a tale of evaporation}


\author{James E. Owen}
\affil{Canadian Institute for Theoretical Astrophysics, 60 St. George St., Toronto, M5S 3H8, Canada}
\email{jowen@cita.utoronto.ca}
\and
\author{Yanqin Wu}
\affil{Department of Astronomy and Astrophysics, University of Toronto, ON M5S 3H4, Canada}
\email{wu@astro.utoronto.ca}

\begin{abstract}
  Inspired by the {\it Kepler} mission's planet discoveries, we consider the thermal
  contraction of planets close to their parent star, under the
  influence of  evaporation.  The mass-loss rates are
  based on hydrodynamic models of evaporation that include both X-ray and EUV irradiation.  We find that only
  low-mass planets with hydrogen envelopes are significantly affected
  by evaporation, with evaporation being able to remove massive
  hydrogen envelopes inward of $\sim 0.1$ AU for Neptune-mass objects,
  while being negligible for Jupiter-mass objects. Moreover, most of
  the evaporation occurs in the first $100$ Myr of the star's lives
  when it is more choromospherically active. We construct a theoretical population of planets with varying core
  masses, envelope masses, orbital separations, and stellar spectral types,
  and compare these against the sizes and densities measured for
  low-mass planets, both in the {\it Kepler} mission and from radial
  velocity surveys. This exercise leads us to conclude that
  evaporation is the driving force of evolution for close-in {\it
    Kepler} planets. In fact, some $50\%$ of the {\it Kepler} planet
  candidates may have been significantly eroded.  Evaporation explains
  two striking correlations observed in these objects: a lack of large
  radius/low density planets close to the stars, and a {\bc possible} bimodal
  distribution in planet sizes with a deficit of planets around
  $2R_\oplus$. Planets that have experienced high X-ray exposures are generally
  smaller than this size, and those with lower X-ray exposures are
  typically larger. {\bc A bimodal planet size distribution is 
  naturally predicted by evaporation model}, where, depending on their
  X-ray exposure, close-in planets can either hold on to hydrogen
  envelopes {\y$\sim 0.5$-$1\%$} in mass, or be stripped entirely.
  To quantitatively reproduce the observed features, we argue that not
  only do low-mass {\it Kepler} planets need to be made of rocky cores overlaid
  with hydrogen envelopes, but few of them should have initial masses
  above $20 M_\oplus$, and the majority of them should have core
  masses of a few Earth masses.

\end{abstract}



\section{Introduction}

The spectacular success of the {\it Kepler} mission has yielded
thousands of planetary candidates
\citep[e.g.][]{Borucki,Batalhaetal12}. Most of these exoplanets are
smaller than Neptune ($4 R_\oplus$) and are likely to have masses
between a few, to tens of earth masses. We can now hope to make
substantial progress in understanding the origin and evolution of
planetary systems by studying the interior composition and orbital
structure of these objects.

However, well-known degeneracies prevents us from resolving their
interior composition. A typical {\it Kepler} planet is likely composed
of a dense core and a tenuous envelope: the core can be {\y 
  volatile}-rich, rock-rich or iron-rich; the envelope can consist of
steam or hydrogen/helium. Unlike the case of main-sequence stars,
radius measurement of these systems (all that is possible for most
{\it Kepler} candidates) can not be used to constrain their
structure. Even in the case where planetary masses are measured (via
transit-timing-variations, TTV, or radial velocity, RV), multiple
solutions for the interior structure exist. For instance, a
Neptune-massed planet, with a density of $2$ g cm$^{-3}$, can have
either a thin hydrogen/helium envelope surrounding a dense rocky core,
or be entirely ice/water dominated
\citep[e.g.][]{degeneracy,rogers_seager}.


But this degeneracy can be broken by studying {\y  a population of
  planets and investigating} how planetary sizes and densities
correlate with their physical environment \citep{WuLithwick}. In
particular, for those planets at close separations from their parent
star ($<$0.2 AU), the total received high energy irradiation over Gyr
time-scales can represent a considerable fraction of their
gravitational binding-energy
\citep[e.g.][]{lammer2003,lecavelier2007,davis2009}.  If these planets
have hydrogen-rich envelopes, evaporation of these envelopes could
markedly reduce their sizes and increase their bulk densities, when
compared to planets further out.

Two recent studies have looked for {\y correlations with separation}.  \citet{ciardi2013} note that, for pairs
of {\it Kepler} planets that are bigger than Neptune ($4 R_\oplus$,
these almost certainly have hydrogen envelopes), the inner planets are
more frequently the smaller ones. However, they do not observe such a
hierarchy in pairs smaller than Neptune. Therefore, they concluded
that the correlations may be a result of the planet formation process,
as opposed to any post-formation process.  This conclusion contrasts
with that arrived by \citet{WuLithwick}; in this study, they measure
planetary masses in a sample of TTV pairs and find that, within pairs,
the inner planets tend to be denser. Moreover, when considering all
available mass measurements together, they find that planet densities
increase with decreasing orbital periods. Therefore, they conclude
that the low-mass planets observed by {\it Kepler} are composed of
dense rocky cores overlaid with various amounts of hydrogen in their
envelopes, which are then sculpted by evaporation. The same density and radius trends are encapsulated in many
of the {\it Kepler} multi-planet systems, e.g., Kepler-11
\citep{Lissaueretal11}, Kepler-18 \citep{kepler18}, Kepler-20
\citep{kepler20} \& Kepler-36 \citep{kepler36}, {\y  as well as
  reported by recent radial velocity studies \citep{Weiss}. Thus an important question has arisien: are the
  observed correlations (radius/density vs. distance) results of formation or evaporation? }

On the theoretical side, there is a large body of literature
discussing planet evaporation, ever since the first hot Jupiters were
discovered
\citep[e.g.][]{lammer2003,yelle2004,tian2005,hubbard2,hubbard1,murrayclay2009,koskinen2010b,owen}. Most
of these deal with gas-giants, and only a few focus specifically on
low-mass planets. We summarize two works here that are of direct
relevance to our discussion.

\citet{owen} develop realistic models of hydrodynamic evaporation for
low-mass planets, including both X-ray and EUV irradiation. They
perform the calculations on planet models of different densities and
radii. They identify that only low mass-planets ($M_p~<60$ M$_\oplus$)
could experience a significant effect from evaporation and suggest
that evaporation may lead to a stability boundary below which all
observed planets should lie \citep[e.g.][]{koskinen}.  Lacking a
thermal evolution model, they could not make predictions on final
planet radius and density. The detailed evaporation models allow
\citet{owen} to calibrate the evaporation efficiency and they argue
that this parameter can vary by orders of magnitude over the stellar
and planetary lifetimes. Their work also highlight the relative
importances of X-ray versus EUV evaporation.

\citet{lopez} provide the only significant attempt to study the
combined effects of thermal evolution and evaporation. Adopting a
simplified energy-limited formalism for evaporation and a constant
efficiency factor -the caveats of which are discussed in
Section~\ref{sec:evap}- and they also infer the presence of an
evaporation threshold. They find that since the evaporation time-scale
scales with planet mass $\times$ planet density, that planets with
mass$\times$density above a critical value would be evaporated 
 {\y  above} it. But as is demonstrated by \citet{owen}, a
simplistic evaporation model may severely over- or under-estimate the
mass-loss rates, especially in the first stages of the stars' and
planets' lives.
%

Now with the large {\it Kepler} data set, evaporation of low-mass
planets 
deserves a better study and a direct comparison with the
observations. Moreover, it is useful to be able to predict the final
planet size and density, for a given initial model. This allows us to
backtrack the original planetary structures at formation. To
accomplish these goals, we need to improve on previous work, in
particular, we need to couple realistic calculations of evaporation
with thermal evolution models.


We describe our approach in Section~3, after briefly reviewing the
evaporation theory in Section~2.  We then compare our theoretical
results directly against observations in Section~4. Our evaporation
theory successfully explains a number of observed facts, establishing
that evaporation is the driving process of evolution for close-in {\it
  Kepler} planets. In fact, some $50\%$ of the currently known {\it
  Kepler} planets may have been significantly affected by
evaporation. Finally, we discuss the caveat and implications of our work in
Section~5 and conclude in Section~6.




\section{Physics of Planetary Evaporation}\label{sec:evap}

Planet evaporation can take place through a variety of mechanisms:
non-thermal escape; thermal Jean's escape and hydrodynamic escape.
Hydrodynamic evaporation occurs when the density of the heated region
is sufficiently high so that the gas is collisional even in the super-sonic region
of the flow. Thus, it is only hydrodynamic evaporation that can
produce high enough mass-loss to affect the structure and sculpt the
planet, which occurs at high incident fluxes; we focus on this
mechanism.

As high-energy photons from the star ionize gas in the upper envelope
(either hydrogen or metals), the newly freed electrons heat up the
local gas and the atmosphere expands. A flow may be initiated that
eventually escapes from the gravitational well of the planet. Let the
efficiency of converting received energy to $P{\rm d}V$ work be
$\eta$, so the mass-loss rate is simply:
\begin{equation}\label{eqn:EL1}
\dot{M}=\eta \frac{L_{\rm HE} R_p^3}{4GM_pa^2}
\end{equation} 
where $L_{\rm HE}$ is the high energy luminosity (X-rays or EUV),
$M_p$ is the planet mass, $R_p$ is the planet radius and $a$ is the
separation from the parent star.

The usual so-called `energy-limited' approach is equivalent of taking
$\eta$ to be an order-unity constant
\citep[e.g.][]{watson1981,lammer2003,lecavelier2007,erkaev2007}.
Under such an assumption, the evaporative time-scale is $M_p/{\dot M}
\propto M_p\rho/F_{HE}$, where $F_{\rm HE}$ is the high energy
flux. So Equation~\ref{eqn:EL1} lends itself to an
evaporation threshold in terms of mass$\times$density \citep{lopez}.

However, the evaporation efficiency is not always constant but may
vary significantly with planet mass, radius and the ionizing flux (see
Section~5). \citet{murrayclay2009} demonstrated that, in the case of EUV
evaporation of hot Jupiters, the `energy limited' approximation is
only valid at low fluxes. At high fluxes, the mass-loss process is
controlled by the ionization/recombination balance, yielding mass-loss
rates that scale as $L_{\rm EUV}^{1/2}/a$, or, the `efficiency'
($\eta$) decreases with flux. This limit (the
`recombination-dominated' regime) is similar to EUV driven evaporation
of gas clumps \citep{bertoldi1990} and protoplanetary discs
\citep[e.g.][]{johnstone1998,hollenbach2000}.

For X-ray ionization, \citet{owen} showed similarly that the
evaporation is not `energy-limited', but rather the `efficiency' is
controlled by line cooling, and is a strong function of planet
mass. They found that cooling is most important for high mass Jovian
planets, since the higher planet escape temperature and larger
physical size mean flow time-scale is long. Then there is then much
time to radiate away the received heating when the flow is still
sub-sonic. For lower mass, Neptune-like planets, the escape velocity
is much lower and the physical sizes smaller, meaning the flow
time-scale is shorter, so much less energy is radiated away and the
corresponding `efficiency' ($\eta$) is higher. While \citet{owen}
found that the X-ray evaporation gives a similar flux-scaling as given
in Equation \ref{eqn:EL1}, the flow is not close to being
`energy-limited' (where $P{\rm d}V$ work is the dominant energy loss
channel), and only begins to approach an energy-limited case at low
masses $<3$ M$_\oplus$. Furthermore, Owen \& Jackson (2012) noted that unlike
the EUV case, since the X-ray driven mass-loss scaled as $L_X/a^2$ at
high and low fluxes there is no transition from line-cooling limited
to energy limited at low X-ray fluxes.

The ionizing flux from a main-sequence star varies by orders of
magnitude through its life. In addition, the size of a planet also
changes over time, under both thermal contraction and mass-loss, along
with the mass of the planet decreasing.  With $\eta$ being a function
of planet mass and radius, as well as the ionizing flux (see
Figure~\ref{fig:eff}), one should not adopt a constant $\eta$ (as in
Lopez et al. 2012) in tracing out the evaporation history of the
planet.  It is important that the correct prescription of evaporation
is used to follow the planetary evolution, particularly for low mass
planets where evaporation can significantly affect the evolution. This
is essential if one wants to make inferences about the initial state
of these planets.

The last issue worth attention is the nature of the ionizing flux.
Which source of high energy luminosity is driving the evaporation has
only been tackled recently. Some authors have purely used EUV flux in
their studies \citep[e.g.][]{lecavelier2007}, while others have used
only the X-rays \citep{jackson2012}.  \citet{owen} solved the flow
problem including both X-ray and EUV irradiation. They found that the
position of the transition from sub-sonic to super-sonic flow (the
`sonic point'), relative to the respective ionization fronts,
determines whether X-rays or EUV is driving the mass-loss. At a
similar ionizing energy flux, they found that X-ray drives the
mass-loss when the flux is high, and EUV dominates the mass-loss when
the flux is low. For the first $\sim 100$ Myrs of their lives,
main-sequence stars have X-ray luminosities that reach up to $\sim
10^{-3}$ of their bolometric luminosities; this fraction falls off
with time roughly as $1/t$ as the stars age
\citep{guedel2004,ribas2005,jackson2012}. 


\section{Evolutionary models of planetary evaporation}

In order to follow the evolution of a close-in planet, we need to
include the effects of thermal cooling of the planet, irradiation of
its upper atmosphere and mass-loss in the form of hydrodynamic
evaporation. Previously, \citet{lopez} have addressed this problem by
coupling the \citet{fortney2007a} and \citet{nettelmann2011} planetary
structure models to a simplistic energy-limited estimate for the
mass-loss.  We aim to achieve a similar goal here, but we will base
our mass-loss rate on a realistic calculation that include both the
X-ray and EUV radiation \citep{owen}.  We make use of the {\sc mesa}
stellar evolution code \citep{paxton2011,paxton2013} to model the
thermal evolution and couple it to the Owen \& Jackson mass-loss
calculations.

\subsection{Method}

The general purpose {\sc mesa} code provides the frame-work to
simulate planetary structure and evolution. Our planet models have
solid inner cores, experience irradiation (under the two stream
approximation, e.g. \citealt{guillot2010}), and evaporation. {\y The {\sc mesa} equations of state is discussed in detail in Paxton et al. (2011), but in the planetary regime is typically based on the SCVH equation of state (Saumon et al. 1995). The opacity tables are used for the irradiation of the atmosphere are based on an updated verision of the Freedman et al. (2008), as detailed in Paxton et al. (2013), where we adopt an opacity $\kappa_*=4\times10^{-3}$ cm$^2$ g$^{-1}$ to the incoming stellar irradiation suitable for a solar type star (Guillot 2010).} 

To include evaporation, we tabulate the mass-loss rates as functions
of ionizing flux, planet mass and planet radius, using the results in
\citet{owen}.  The tabulated results span a range in planet mass from
$1 M_\oplus$ to $3 M_J$, and in planet radius from $1 R_\oplus$ to
$\sim 0.01$ AU (roughly the Hill radius at $0.1$ AU for our most
massive planet).  For the ionizing X-ray flux, we adopt the observed
{\y  relation between this flux and} stellar spectral type and {\y 
  stellar} age \citep{jackson2012}. In general, the X-ray luminosity
saturates at $\sim 10^{-3}$ of the stellar bolometric luminosity
during the first $\sim 100$ Myrs of the star's life, and decays
approximately as $1/t$ afterwards.  Furthermore, following
\citet{owen}, we take the EUV luminosity to follow the same time
evolution of the X-rays.

Since convective transport is the dominant heat transport mechanism in
the planet's envelope, the thermal structure of the envelope is set to
be initially adiabatic, in the absence of irradiation. The radius of
the planet is defined to be at a radius where the optical depth to the
incoming bolometric radiation is $2/3$, and is typically around
miliBar (for younger and lower mass planets) to Bar (for older and
more massive planets) pressures for the planets considered here. This
planet radius is also taken as the input radius in the evaporation
model.  Since the atmosphere's underlying scale height is typically
small compared to the planet radius, such an approximation is
accurate. {\y  For the same reason, we have ignored the difference
  between the above radius and the radius a planet exhibits at transit
  \citep{Hubbardetal2001}.} But {\y  such a simplification} may
break down for the closest in planets at the earliest times, where
there may be a $\sim 10\%$ difference in a planet's optical
photosphere and the base of the evaporative flow. However, given that
planets quickly contract through such a phase, the effect will be
negligible when integrated over the Gyrs of planetary evolution.

\begin{figure}
\centering
\includegraphics[width=\columnwidth]{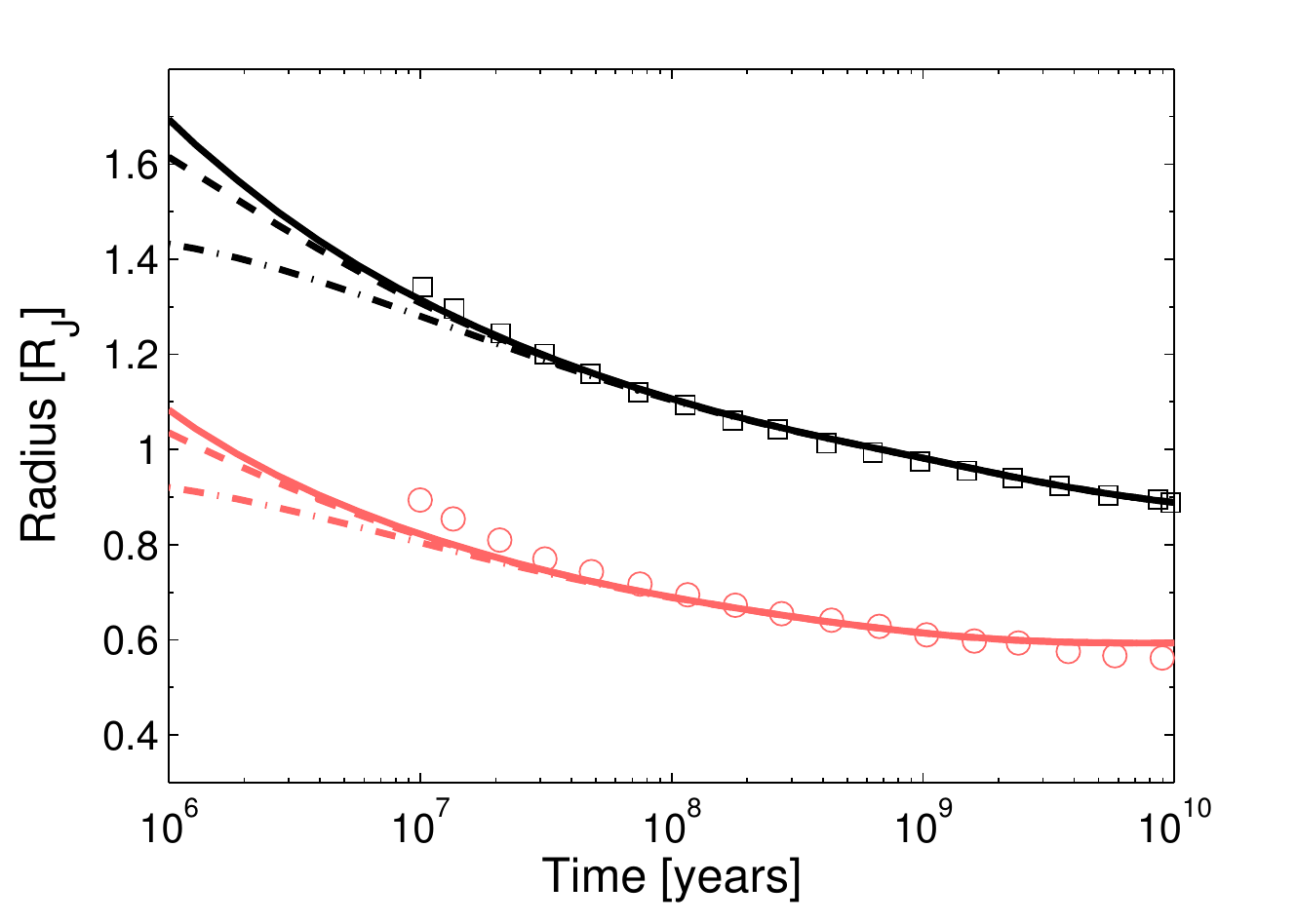}
\caption{Benchmark calculations of our modified version of {\sc mesa}
  against the calculations of Fortney \& Nettelmann (2010) for 95
  (squares) \& 32 (circles) M$_\oplus$ planets at a seperation of
  0.045 AU from a sun-like star. Our calculations are shown for three
  initial cooling times of $10^{5}$ (solid line), $10^{6}$ (dashed
  line) and $10^{7}$ (dot-dashed line) years. Agreement is at the $\le
  5\%$ level in all cases.}\label{fig:benchmark}
\end{figure} 

Our planets are composed of two separate regions: an envelope of
Hydrogen/Helium with a metalicity at solar abundances (as used for the
evaporation models), and a solid core. Given the large range of
possible core compositions, as a starting point we focus on a pure
rock core, using the mass-radius profile provided by
\citet{fortney2007b,fortney2007a}, with core masses ($M_c$) of 6.5,
7.5, 10, 12.5 \& 15 M$_\oplus$ adopted. {\y These cores are assumed to of fixed radius and do not evolve during the planets evolution. However, the cores do have a thermal content from from both radioactive decay  and thermal heat capacity where we adopt an earth-like value (see Nettelmann et al. 2011, Lopez et al. 2012). Both of which are implemented in {\sc mesa} using the core luminosity function (Paxton et al. 2013).}  As current planet formation
models are unable to provide a good handle on the initial thermal
properties of formed planets (e.g. `hot' or `cold' start) we consider
initial models with a wide range of initial radii. This initial radii
(or more correctly entropy) is parametrised in terms of an initial
cooling time ($t_{\rm cool}$), which we define as the ratio of a
planet's initial internal energy to a planet's initial luminosity. We
could of course parametrise the initial entropy in terms of some other
variables; however, the initial cooling time is perhaps the most
interesting as this be compared to protoplanetary discs lifetimes
$\sim$3~Myr
\citep[e.g.][]{haisch2001,hernandez2007,owen2011,armitage2011}. 

{\y In order to make sure our model is following the evolution of low-mass planets accurately we benchmark our calculations against the models
  presented in Fortney \& Nettelmann (2010 - their Figure~10); the
  calculations were performed using the Fortney et al. (2007a) code
  for a 95 \& 32 M$_\oplus$ planet with a 25 M$_\oplus$
  core\footnote{We note the Fortney et al. (2007) calculations used a
   core with no thermal contribution and we do the same for the benchmark calculations.} that
  is a 50/50 mix (by mass) of ice and rock. We follow the evolution of
  these planets under the influence of irradiation by a sun-like star,
  but no evaporation. We find excellent agreement to these
  calculations at the $\le 5\%$ level giving confidence our modified
  version of {\sc mesa} is performing as expected at low-masses. In
  Figure~\ref{fig:benchmark} we show the radius evolution of our
  calculations for cooling times of $10^{5}$ (solid line), $10^{6}$
  (dashed line) and $10^{7}$ (dot-dashed line) years. These
  calculations are compared against the results from the
  \citet{fortney10} at a separation of 0.045AU shown as points for
  both the 95 (squares) \& 32 (circles) M$_\oplus$ planets.}
 
 For our calculations we
choose values of $t_{\rm cool}$ (computed in the absence of
irradiation) in the range $3\times 10^6-10^8$ years to span a range of
`hot' start and `cold' start model, similar to those values chosen by
Lopez et al. (2012). We note that with cooling times $< 3\times
10^{6}$ years most low mass planets at separations $<0.1$ AU have
initial radii larger than their Hill radii, and they cannot be
considered hydrostatically bound objects.

For each cooling time and each core mass, we construct 40 models with
a range of envelope masses from 3 M$_J$ to a few percent of the core
mass, logarithmicly spaced in envelope mass. The planets are then
evolved forward in time, under the influence of evaporation and
irradiation for 10~Gyrs or until the entire envelope is
lost\footnote{Thus, we ignore any subsequent mass-loss from the core
  due to sublimation, which may happen at the highest equilibrium
  temperatures $T_{\rm eq}\ga 2000$ K \citep[e.g.][]{PBC13}.}.  Our
integrations begin at $3$ Myrs, a time at which the protoplanetary
disc clears and the planet is fully exposed to the X-ray and EUV
irradiation. In general the planets evaporation begins in the X-ray
driven phase, and may switch to EUV driven at some late time. Where
the planets' final mass is set by a few 100 Myrs.

\subsection{Jupiter-like planets}\label{sec:jupiter}
\begin{figure}
\centering
\includegraphics[width=\columnwidth]{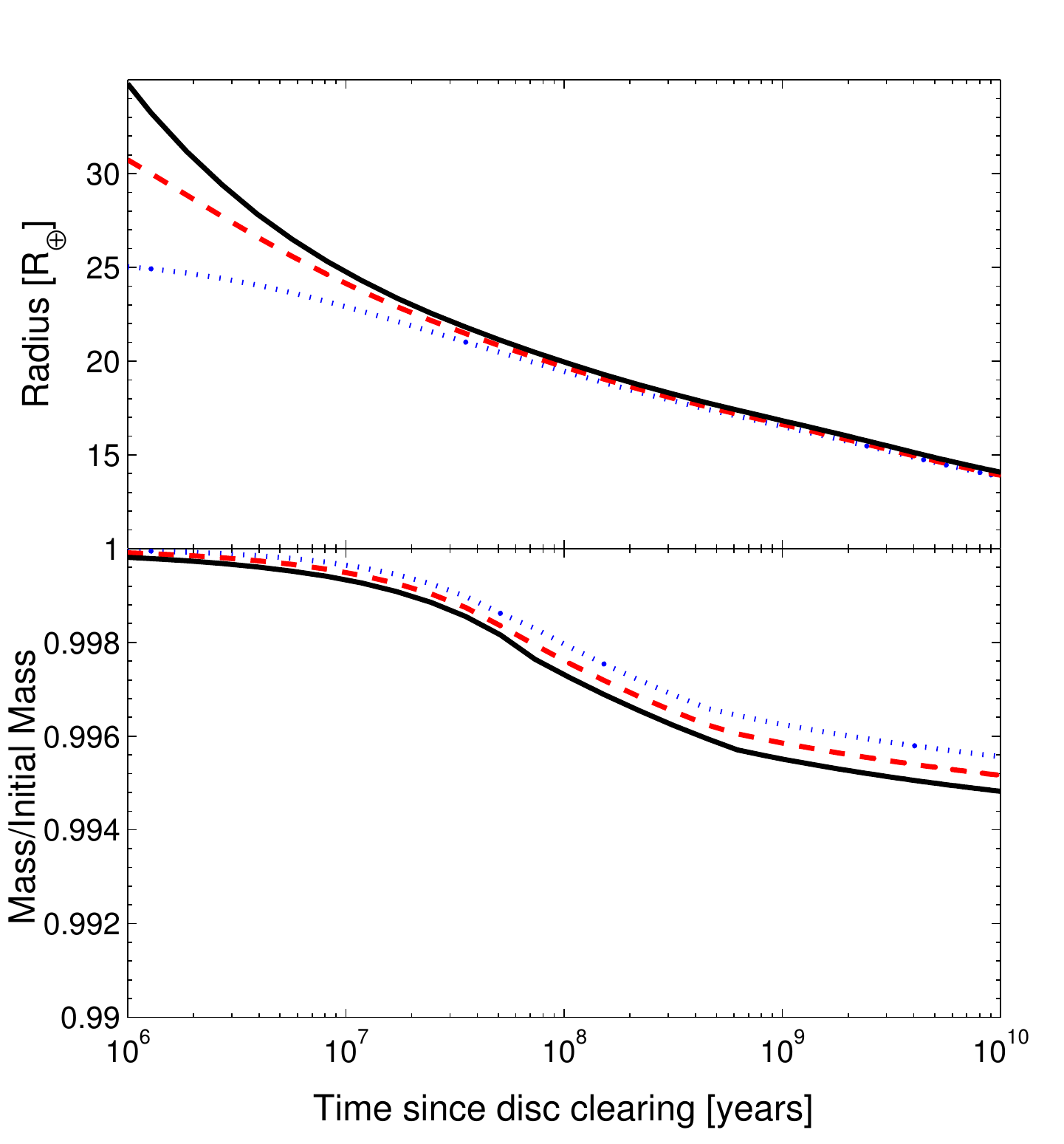}
\caption{The top panel shows the radius evolution of Jupiter like
  planets as a function of time since gas disc clearing, at a very
  close separation to their parent star ($\sim 0.025$ AU); the bottom
  panel shows the mass evolution of this planet. The
  solid line represents a planet with an initially high entropy with an initial
  cooling time of $10^6$ years, the dashed line has an initial cooling
  time of $10^{7}$ years, and the dotted line has an initially low
  entropy with an initial cooling time of $10^{8}$ years.}
\label{fig:Jupiter}
\end{figure}
Jupiter mass planets close to their central stars represent the case
where there is direct observational evidence
\citep[e.g.][]{vidalmadjar2003,vidalmadjar2004} of evaporation
occurring. Thus, it is worth investigating whether their are any
evolutionary consequences for the evaporation of high-mass planets.
For example, \citet{baraffe2004} noted that if the evaporation rate
was high enough, so that the evaporation time ($\sim M_p/\dot{M}$)
became comparable to the thermal time of the planet's envelope, then a
Jupiter mass planet could loose its entire envelope rapidly. Such an
inference was based upon the rather unrealistic assumption of $100\%$
mass-loss efficiency. In reality, at high masses the mass-loss rates
never reach such high values (see Owen \& Jackson 2012 for a more
detailed discussion). To illustrate this we show the evolution of a
Jupiter-like planet at a very close separation of $0.025$ AU around a
solar type star in Figure~\ref{fig:Jupiter}. We plot the radius and
mass evolution of planets with a $15$M$_\oplus$ rock core and cooling
times ranging from the very short ($10^{6}$ years - solid) to the very
long ($10^{8}$ years- dotted). Figure~\ref{fig:Jupiter} clearly shows
that evaporation is unable to effect the planets' evolution and
planets with very different initial cooling times end up on almost
identical evolutionary tracts at late times. With the mass-change in
the planets at the $<1\%$ level, as argued for by \citet{hubbard2}.


Therefore, while Jupiter mass planets currently provided the best
opportunity for actually studying the hydrodynamics of evaporation by
directly probing the flow, they do not provide a good laboratory for
studying the evolutionary consequences of evaporation and we must go
towards lower mass planets where the evolutionary effect will be more
pronounced.
  
\subsection{Low-mass planets}\label{sec:standard}

\begin{figure*}
\centering
\includegraphics[width=0.85\textwidth]{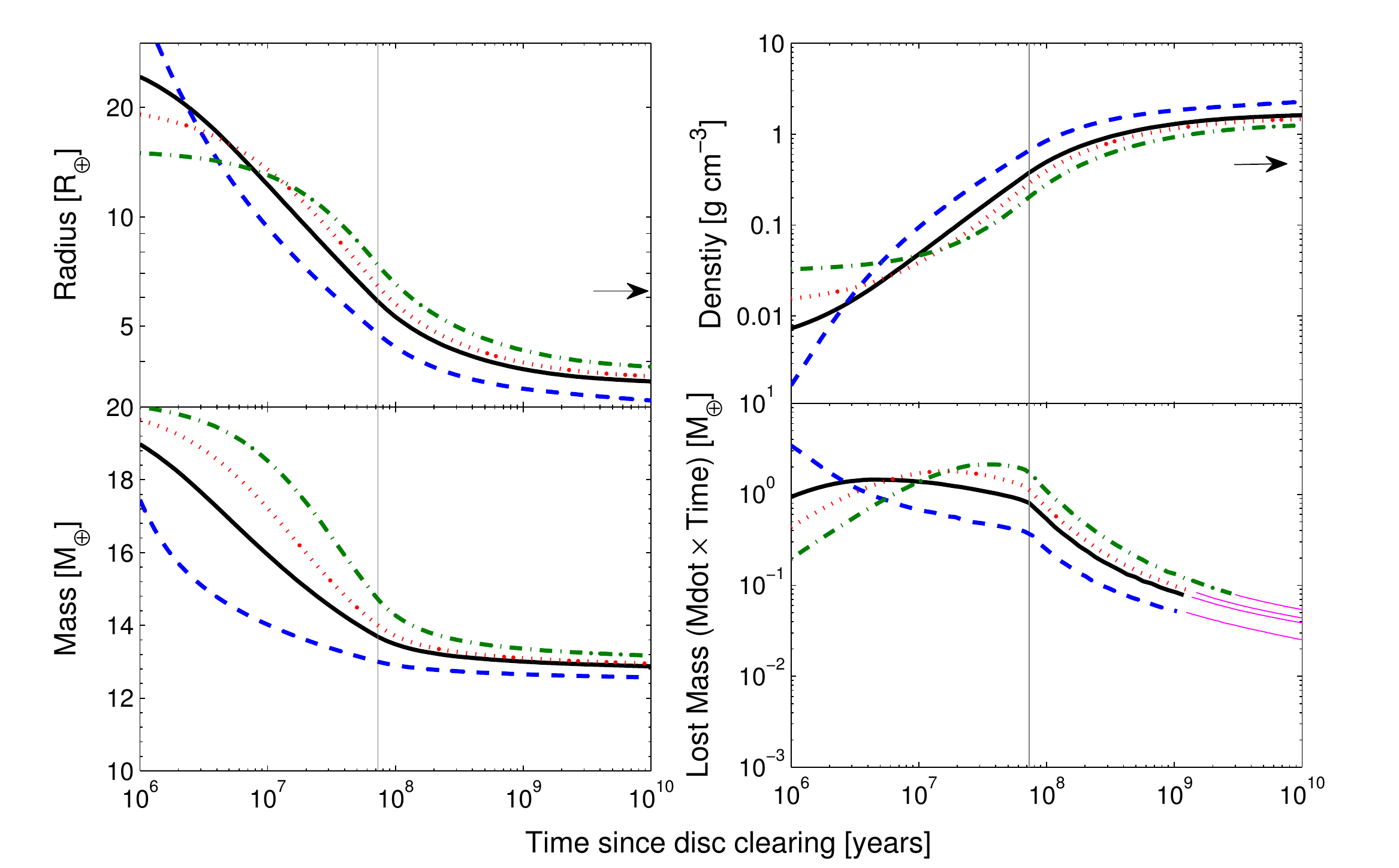}
\caption{The evolution of a 20 M$_\oplus$ planet with a 12.5
  M$_\oplus$ core at a separation of 0.05 AU is plotted, with four
  initial cooling times: $3\times10^6$ years (dashed); $1\times10^7$
  years (solid); $3\times 10^7$ years (dotted); $1\times10^{8}$ years
  (dot-dashed). In the upper left we show the radius evolution, the
  density evolution is shown in the upper right, the mass evolution in
  the bottom left and in the bottom right we show the time evolution
  of $\dot{M}\times{\rm age}$ which indicates the times at which most
  mass-loss is occurring, with the thin continuation to these lines
  show the point at which the evaporation switches from X-ray driven
  to EUV driven.  The final properties of the same
  planet not undergoing evaporation are shown by the small
  arrows.}\label{fig:standard}
\end{figure*}  

Unlike the Jupiter-type planets discussed above, several authors have
suggested that the effects of evaporation will be more prominent
around lower mass planets
\citep[e.g.][]{hubbard2,baraffe2008,jackson2012,owen,lopez}. At lower
masses evaporation can begin to sculpt the planet population, removing
significant amounts of a planet's envelope during its lifetime. To
investigate this we show the evolution of a low mass planet in
Figure~\ref{fig:standard}, where we consider the evolution of a
`standard' model which is an initially 20 M$_\oplus$ planet with a
12.5 M$_\oplus$ core. Figure~\ref{fig:standard} shows the evolution of
the `standard' model for the full range of initial cooling times
considered. The planets are at a close separation of $\sim$0.05 AU and
have equilibrium temperatures of 1300K. Where we define the
equilibrium temperature as the black body temperature at a given
separation, i.e.:
\begin{equation}
T_{\rm eq}=T_*\sqrt{\frac{R_*}{2a}},
\end{equation} 
where $T_*$ and $R_*$ are the parent star's temperature and radius
respectfully.

The panels in Figure~\ref{fig:standard} show the qualitative features
of the evolution of low-mass planets. In particular, planets with
higher initial entropies have initially larger radii and therefore
loose more mass. The result is that planets with shorter initial
cooling times end up with, smaller radii and higher densities than
planets with longer initial cooling times. At low masses evaporation
plays a strong role in planetary evolution, which in the case of our
`standard' model makes the planets a factor of $\sim 2$ smaller and a
factor of $\sim 4$ denser compared to the same planet that is not
undergoing evaporation. In this case an initial envelope containing
$\sim40$\% of the original mass is evaporated down to one containing
only few percent of the total mass, and in the most extreme case (for
the planet with an initial cooling time of $3\times10^{6}$ years)
evaporation leaves a planet with $0.5\%$ of the total mass in a
Hydrogen/Helium envelope.

The bottom, right-hand panel of Figure~\ref{fig:standard} shows the
evolution of $\dot{M}\times{\rm age}$, which indicates when the
mass-loss is most significant. This plot shows that in all cases the
mass-loss is most important at roughly the point where the X-ray
luminosity begins to decline. This is easy to understand as at early
times the planets are large and the fluxes are high, so the planet can
absorb a significant amount of high energy radiation. Once the X-rays
begin to decline the planet evolution is affected less by evaporation
and once the evaporation switches to EUV driven the final planet
properties have already been `frozen' in, similar to the results from
the previous models calculated by Owen \& Jackson (2012) and Lopez et
al. (2012). This is a rather generic feature of all the evolutionary
models, where the saturation time-scale for the X-rays sets the length
of time which evaporation is important in driving the planets'
evolution, and that there is very little change in the planets' mass
from 100 Myrs to 10 Gyrs.


\section{Population study and comparison against observations}
\label{sec:pop}

We have shown that, for low mass planets close to their parent star, a
significant amount of the planet's gaseous envelope, or even an entire
atmosphere, can be removed in Gyr time-scales.\footnote{\y We ignore
  all Jovian planets from now on. When comparing against Kepler data,
  this is naturally achieved by only plotting planets in multiple
  systems.}  Given the prominent role evaporation can play, we can use
this evolution to make inferences about the initial structure of
observed close-in planets, and provide clues as to their formation.
In the following, we provide comparisons between our theoretical
models and observations, mostly using data from the {\it Kepler}
mission. The effects of evaporation are clearly visible in data. In
fact, the evaporation model naturally explains a number of remarkable
correlations seen in the {\it Kepler} data.

\begin{figure}
\centering
\includegraphics[width=\columnwidth]{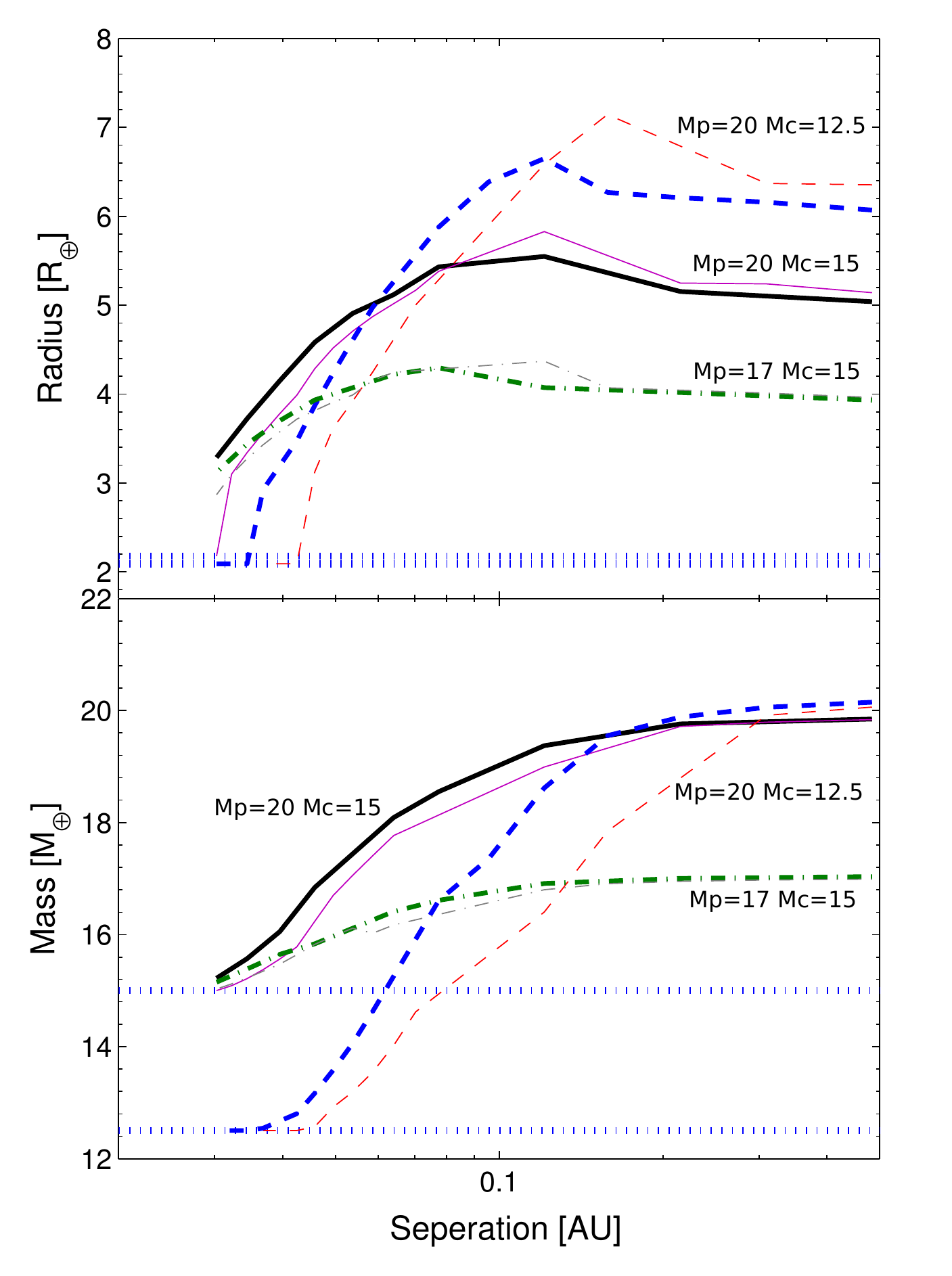}
\caption{Final planet radius (top panel) and mass (top panel) as a
  function of separation, for models with different initial core mass
  (labelled as $M_c$), total mass (labelled as $M_p$) and initial
  cooling time (thin lines for $3\times 10^6$ yrs and thick lines for
  $10^8$ yrs, longer cooling time indicates lower internal
  entropy). The central star is sun-like.  The straight dotted lines
  show the masses and radii of the rocky cores alone.  Planets inward
  of $\sim 0.1$ AU experience significant mass-loss and can retain
  only a fraction of their primordial atmospheres (depending on the
  core mass), with the hottest ones exposed to bare cores. As a
  result, the maximum radii of these planets decrease inward.}
\label{fig:prop_sep}
\end{figure}


In Figure~\ref{fig:prop_sep}, we plot the final planet mass and radius
as a function of separation for planets with initial properties which
vary from our `standard' model, where central star has 1 M$_\odot$ in
all cases.  At a given planet mass, a lower core mass results in
initially larger planets, which means that the planet can absorb a
larger fraction of the X-ray flux, driving stronger evaporative flows.
So planets with smaller cores may lose the entire envelope at a larger
separations than those with larger cores (see Figure
\ref{fig:rad_sep}). The initial cooling time shows the same effect as
discussed above, with shorter cooling times resulting in more
mass-loss due to the initially larger radii; however, this effect is
much less important than the effect of initial core mass.  Finally,
evaporation can drive convergent evolution. For instance, planets at
$< 0.05AU$ that had initial 15 M$_\oplus$ cores, but envelopes of 2 \&
5 M$_\oplus$ end with somewhat identical structures This means that,
at least in some cases, it may be difficult to retrieve the planets'
initial structure, particularly at low envelope mass-fractions.


\subsection{Upper Envelope in Planet Radius}
\label{subsec:upper}
 
The {\it Kepler} transiting-planet catalogue is the most extensive
catalogue of planets on close orbits to their parent star; for better
statistics, we use the {\it Kepler} object of interest (KOI)
catalogue, which lists the radii (not mass) of planet candidates.
Most of the KOIs have not been confirmed as planets, and there is a
certain, but low, percentage of false positives \citep{morton11}. To
minimize contamination, we choose to consider only KOIs that have been
identified as being in multiple systems, where the significance of
planetary nature is considerably higher ($\ga$ 95\%
\citealt{lissauer2012}). {\bc The use of only multi-planet systems,
  while the most robust, may introduce some implicit biases. In
  particular, planets that may have been dynamically moved to small
  orbital periods at late times, will have a undergone a different
  evolutionary path. Since we are interested in the long term
  evolution of planets due to evaporation, then the multi-planets KOIs
  represents the cleanest sample to begin with.}

Plotting planet radius against separation for planets in multi-planet
systems around solar-type stars ($T_*=5200-6200$ K) we notice an
upper-envelope in planet size that rises with separation.  This has
also been noticed by \citet{ciardi2013} and \citet{WuLithwick}.  The
lack of large planets at small separations is statistically
significant: we cut the sample into 4 regions: large and small planets
with a division at $R_p = 5$ R$_\oplus$ and hot and cold planets with
a division at $T_{\rm eq}=1000$K, then comparing the ratio of large
cold planets to the ratio of small cold planets. Where one would
expect to observe $\sim 20$ hot large planets if these are the same
populations, we find only $3$.  

\begin{figure}
\centering
\includegraphics[width=\columnwidth]{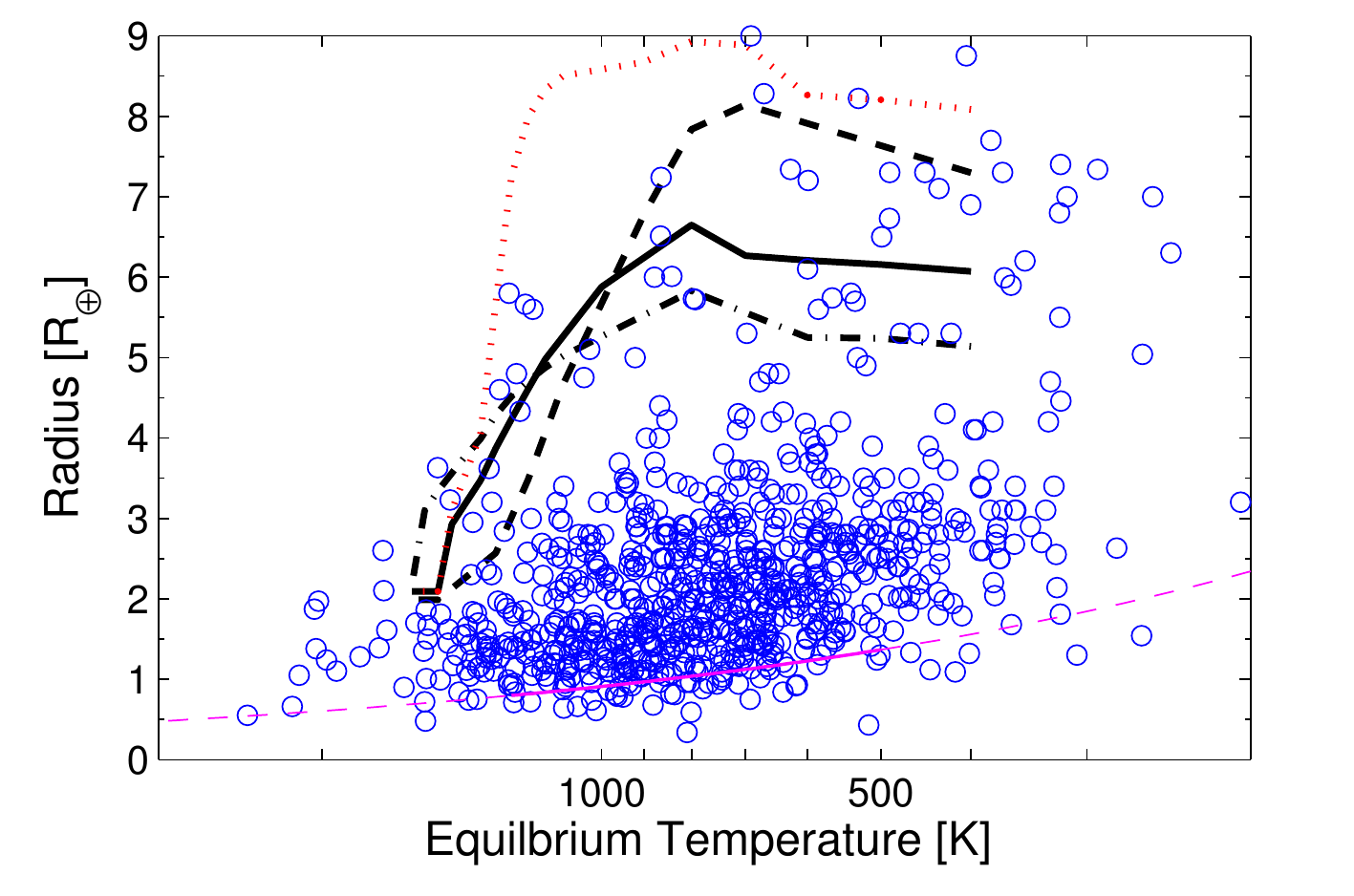}
\caption{The upper envelope of planet sizes as a function of
  equilibrium temperature . The open circles show the Kepler Object of
  Interests that are around solar type stars ($T_*=5200-6200$ K) and
  are in multiple transiting systems. The black curves are the
  theoretical final radii for planet models with initial mass 20
  M$_\oplus$ and 10 (dashed), 12.5 (solid) \& 15 M$_\oplus$
  (dot-dashed) of rocky cores. The dotted line are for $\sim 30$
  M$_\oplus$ planets with 12.5 M$_\oplus$ cores. All models here have
  an initial cooling time of $10^{7}$ years. {\y The thin line shows
    the 50\% completeness limit calculated by \citet{Petigura} -solid-
    and extrapolated -dashed- to larger and smaller separations.}}
\label{fig:envelope}
\end{figure}

{\y We also argue that selection effect in the {\it Kepler} pipeline
  would not produce such an upper envelope. KOI candidates have to
  have a certain signal to noise ratio to be identified. This favours
  detecting smaller planets closer to the stars, as they have produced
  a larger number of transits during the mission lifetime. The
  completeness radius (sizes above which the KOI catalogue is
  complete) will rise with planet periods, similar to the upper
  envelope discussed here. We draw such a curve for the sub-sample
  considered by \citep{Petigura} in Figure~\ref{fig:envelope}, which
  lies far below the upper envelope. Though more noisy stars than
  those considered by \citep{Petigura} will move this curve upward, it
  will not explain the upper envelope.}

This upper envelope is naturally
explained by evaporation: low density planets can not survive in the
environment of high ionizing flux. Quantitatively, evaporation of $20
M_\oplus$ planets with rocky cores of masses 10-15 $M_\oplus$ provides
a good fit to the upper envelope, as is shown in Figure
\ref{fig:envelope}.

 There is another implication to this agreement. If there were a
  significant population of planets with initial masses higher than,
  say, $30 M_\oplus$ (dotted line in Figure~\ref{fig:envelope}), we would not
  expect to observe the same upper envelope. These more massive
  planets can hold on to their atmospheres, much like the hot Jupiters
  can (see Section~\ref{sec:jupiter}), and would populate the upper left
  region in  Figure~\ref{fig:envelope}. The absence of these more massive planets is
  interesting: perhaps not coincidentally, as $20-30$ M$_\oplus$ roughly
  corresponds to gap-opening mass in this region, suggesting that
  there is ceiling to how much gas the planets can accrete in this
  neighbourhood. The core mass also somewhat affects the upper
  envelope. We find that models with roughly half of the total mass in
  the rocky cores best reproduce the data, consistent with the
  conclusion in \citet{WuLithwick}.   In contrast, the planet's initial
cooling time makes
  little difference to the results.

\subsubsection{Dependence on Stellar Spectral Types}
\label{sec:mass_eff}
The mass of the parent star is an important consideration. The parent
stars direct influences on the evaporation, through its gravity is
small.  However, the total received X-ray flux at a fixed bolometric
flux (so fixed equilibrium temperature) can vary greatly, with
late-type stars being significantly more X-ray luminous when
integrated over Gyr time-scales
\citep{guedel2004,guedel2007,jackson2012}. We demonstrate this by
considering the evolution of the `standard' planet discussed above
($M_p=20$ M$_\oplus$, $M_c=12.5$ M$_\oplus$), at a fixed equilibrium
temperature (1300K) around a later-type (0.5 M$_\odot$) and
earlier-type (1.5 M$_\odot$) star compared to a planet around a solar
type star, all with initial cooling times of $10^{7}$ years.

\begin{figure}
\centering
\includegraphics[width=\columnwidth]{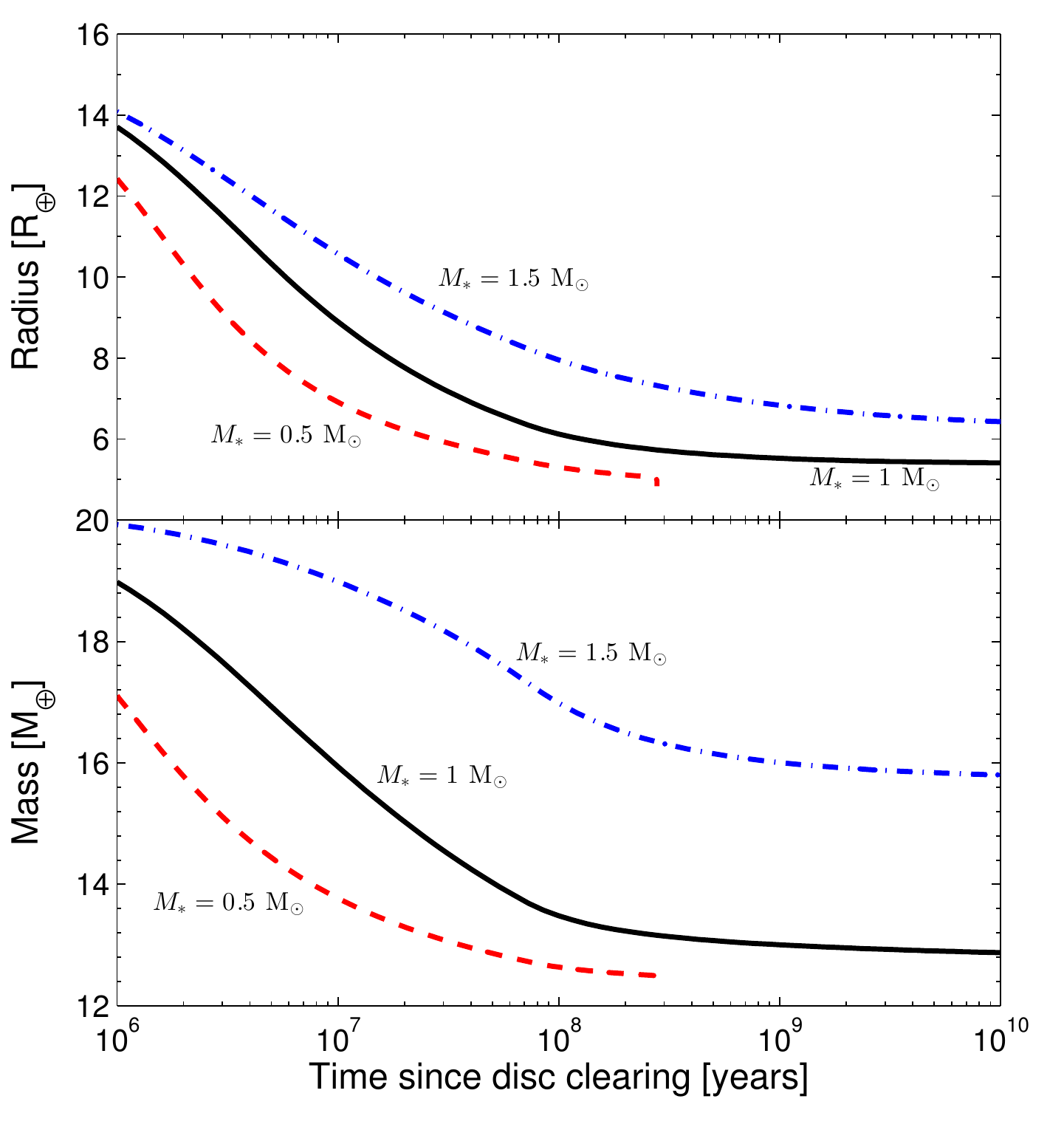}
\caption{The panels show the radius (top panel) and mass (bottom
  panel) evolution of a planet (20 M$_\oplus$) with an equilibrium
  temperature of 1300K ($\sim$0.05 AU around
  a solar type star) and $12.5$ M$_\oplus$ rock core, around a 1.5
  $M_\odot$ star (dot-dashed), 1 $M_\odot$ star (solid) \&
  0.5$M_\odot$ star (dashed). The solid lines are identical to the
  solid lines in Figure~\ref{fig:standard}. }
\label{fig:star_mass}
\end{figure}

The evolution of these planets, is shown in
Figure~\ref{fig:star_mass}, where the radius evolution is in the top
panel and the mass evolution in the bottom panel.

 Naively, one would expect a similar evolution as the
bolometric flux revived is identical in all cases. However, the
variation of X-ray luminosity with stellar mass results in
qualitatively different evolutionary paths for the planets, with
order-unity differences in both the final planet mass and final planet
radius.  The planet around the 0.5 M$_\odot$ star has had its envelope
completely removed, whereas the planet around the 1.5 M$_\odot$ planet
still has a $\sim 3$ M$_\oplus$ envelope remaining after 10 Gyrs.

We can go further and compare our evaporation threshold found above
for solar type stars to KOIs around other types of host stars. Lower
mass stars (e.g., M-dwarfs) have higher X-ray flux, when compared to
their bolometric luminosities. They also remain choromospherically
active for longer periods. Figure \ref{fig:star_mass} shows that,
indeed, at the same equilibrium temperature (measuring the bolometric
flux),  {\y the upper envelope around lower mass stars appear
  to be suppressed to smaller planet sizes.}

\begin{figure}
\centering
\includegraphics[trim=65 135 50 100,clip,width=\columnwidth]{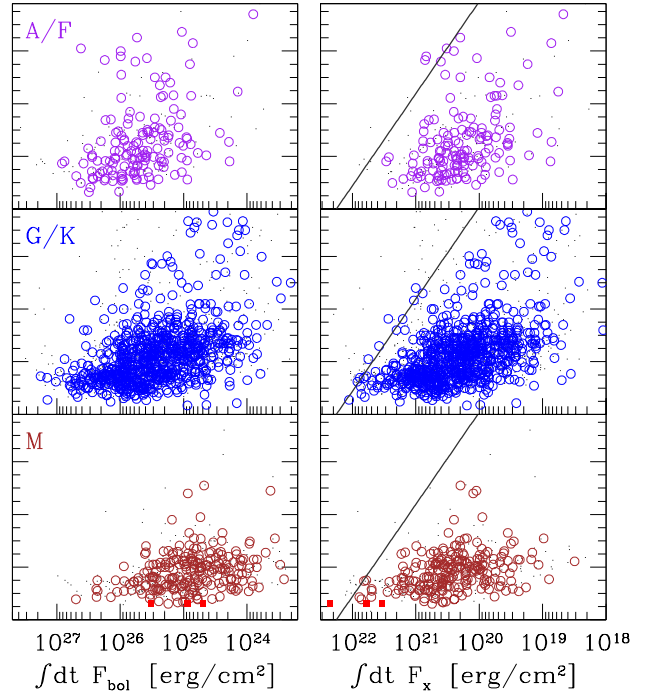}
\caption{The observed radii of KOIs are plotted against their
  bolometric exposure (bolometric flux received at surface integrated
  over 10 Gyrs, left panels), and X-ray exposure (same but for X-ray
  exposure, using \citealt{jackson2012} values, right panels). The
  objects are roughly separated into those around A/F stars (top
  panels), G/K stars (middle-panels) and M-dwarfs (lower panels, the
  red squares stand for the system KOI 961).
  KOI multis are shown as open circles while {\bc KOI singles as small
  dots}.  As predicted by our models of x-ray evaporation, planets
  around stars of different spectral types, while having very
  different bolometric exposure, have roughly the same maximum sizes
  at a given x-ray exposure. The solid lines show the theoretical
  evaporation threshold derived from a linear fit to the black
  evaporation curves shown in Figure~\ref{fig:envelope}. The right
  panels demonstrate that planets around different types of stars
  satisfy the same maximum radius/X-ray exposure relation as that
  found around solar-type stars. One cannot perform the same exercise in the left panels.}
\label{fig:wu}
\end{figure}

{\y Currently, the numbers of candidates around A/F and M stars are
  not as large as those around G/K stars. Thus, a fully quantitative comparison is not possible at this stage. Moreover, planet radii determination
  around M-stars suffer large uncertainties
  \citep{Muirhead,Mann,morton}, and radius determination for hot stars
  can also be polluted by the presence of sub-giants \citep{KIC}, and must bare this in mind when drawing inferences.} We
determine the theoretical evaporation threshold by extracting a linear
relation between the maximum radii and equilibrium temperature, for
planets with an initial mass of 20 M$_\oplus$ and a core mass in the
range $10-15$ M$_\oplus$, that are orbiting around a solar-type star
and have equilibrium temperature in the range of $500-2000$ K.  Since
the total mass loss roughly scales linearly with the integrated X-ray
flux, we expect the same radius threshold to apply to planets around
all spectral types, when we arrange them by their X-ray exposure.
This is shown in Figure~\ref{fig:wu}, where the planets are separated
by the spectral type of their parent star.\footnote{Since the
  evolution of the X-ray luminosity is poorly known for stars $<0.45$
  M$_\odot$, we approximate the X-ray evolution of these stars with
  that of a 0.45 M$_\odot$ star.} In contrast, we show the same planet
radii plotted against their bolometric exposure.  Planets satisfy the
same evaporation threshold only when one considers X-ray
exposure. This argues that the ionizing flux, not the bolometric flux,
is what determines the upper envelope. {\bc Furthermore, we also show the KOIs in single planet systems in Figure~\ref{fig:wu} as small filled circles, which show the same behaviour as the multi-planet KOIs, although with slightly more scatter.}

\subsection{Distribution of Radius}

\begin{figure}
\centering
\includegraphics[width=\columnwidth]{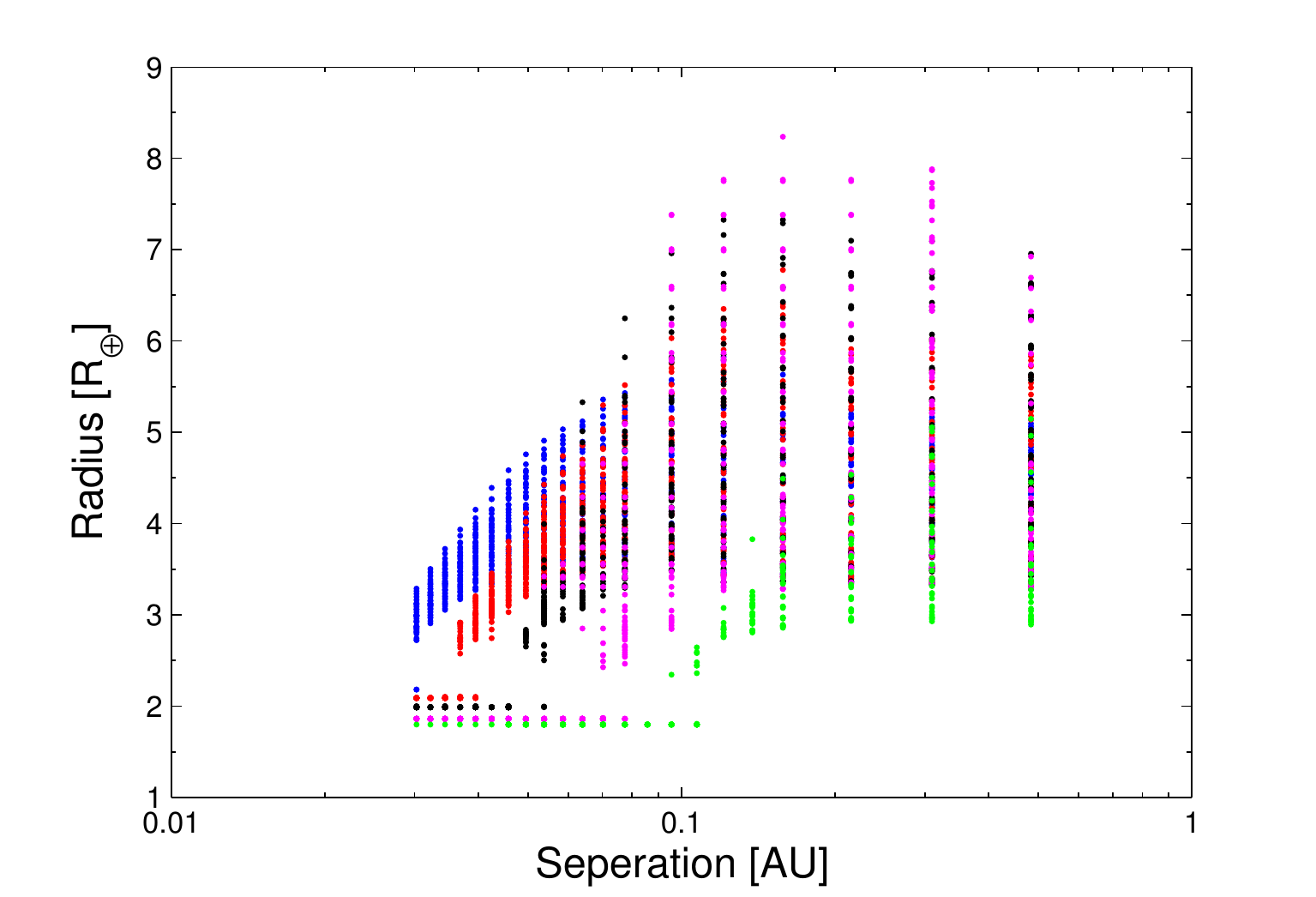}
\caption{  The final   radii
  of planets with an initial mass $<20$M$_\oplus$, as a function of
  separation from a sun-like star. Different colours
  represent different core masses with
  blue standing for 15 M$_\oplus$ core, red for 12.5 M$_\oplus$ core,
  black 10 M$_\oplus$, magenta 7.5 M$_\oplus$ \& green 6.5 M$_\oplus$.
  We consider atmospheres as low in mass as one percent of the
    core mass.  For all models, there is a
    critical separation within which the planets are evaporated down
    to bare cores. This separation is larger for models with smaller
    core masses.  }
\label{fig:rad_sep}
\end{figure}

Our analysis in the previous sections suggest that the observed radius
cut-off in KOIs is related to the fact that the most massive
KOIs\footnote{\y Aagin, we exclude Jovian planets from this
  discussion.} have masses not much exceeding $20 M_\odot$ and that
their core masses are roughly half of their total masses. Here, we
investigate the nature of all KOIs by studying the overall radius
distribution.

 In Figure \ref{fig:rad_sep}, we present the final radii of multiple
  sequences of planet models with initial cooling times in the range $3\times10^{6}-10^{8}$ years. These have core masses from 6.5
  $M_\oplus$ to 15 $M_\oplus$, and atmosphere masses from approximately one percent
  of the core mass to much larger values. The total
  planet masses are restricted to $< 20 M_\oplus$. We do not consider
  atmospheres with masses below one percent, motivated by the
  discussion below.  The radii are evaluated after 10 Gyrs of orbiting
  around a sun-like star, though the values differ little if we
  instead evaluate at 1 Gyrs
  \citep[e.g.][]{lopez,lammer2013}. Figure \ref{fig:den_sep} shows the
  corresponding planet densities as a function of separation and is
  discussed in Section~\ref{subsec:density}.  
%

  The overall population show the general feature noted previously,
  that the radius decreases with decreasing separation.  One
  particularly interesting feature that appears is a gap in radius
  between planets that have gaseous envelopes, and those that are bare
  cores.  For instance, for the 6.5 $M_\oplus$ core models, inward of
  $\sim 0.1$ AU, all planets have their atmospheres stripped away with
  their final sizes reflecting that of their naked rocky cores; while
  outside $\sim0.1$ AU, planets can retain atmospheres that are at
  least a fraction of a percent in mass, consequently they have sizes
  that are markedly larger.  This bifurcation generates a gap in
  planet radius. The orbital separation at which this gap appears is
  smaller for planets that have bigger cores and stronger surface
  gravity. However, inside $\sim 0.03 AU$ there are no surviving
  planets with gaseous envelopes.

The origin for this gap is easy to understand. As planets lose their
hydrogen atmosphere, they become increasingly compact and dense, which
reduces the mass-loss rate. However, there is also less hydrogen to
lose. So for planets inside the critical separation, the loss is
total (e.g. Baraffe et al. 2006; Lopez et al. 2012). While for planets just outside the critical separation, there
is a minimum atmosphere mass the planets need to avoid complete
stripping. Any thinner atmosphere will be easily eroded. This mass is
roughly $1\%$ of the total mass and corresponds to roughly an order
unity modification to the planet radius. So we expect to see a gap in
planet size. {\y Such a gap may become less pronounced when different core compositions are considered, but the basic property that small atmospheres are unstable to complete evaporation will always result in a region where planets are unlikely to end up. Observationally determining the presence of such a gap will place strong constraints on the model and further characterising the gap will allow useful inferences about the dominant core mass/composition.}

\begin{figure}
\centering
\includegraphics[width=\columnwidth]{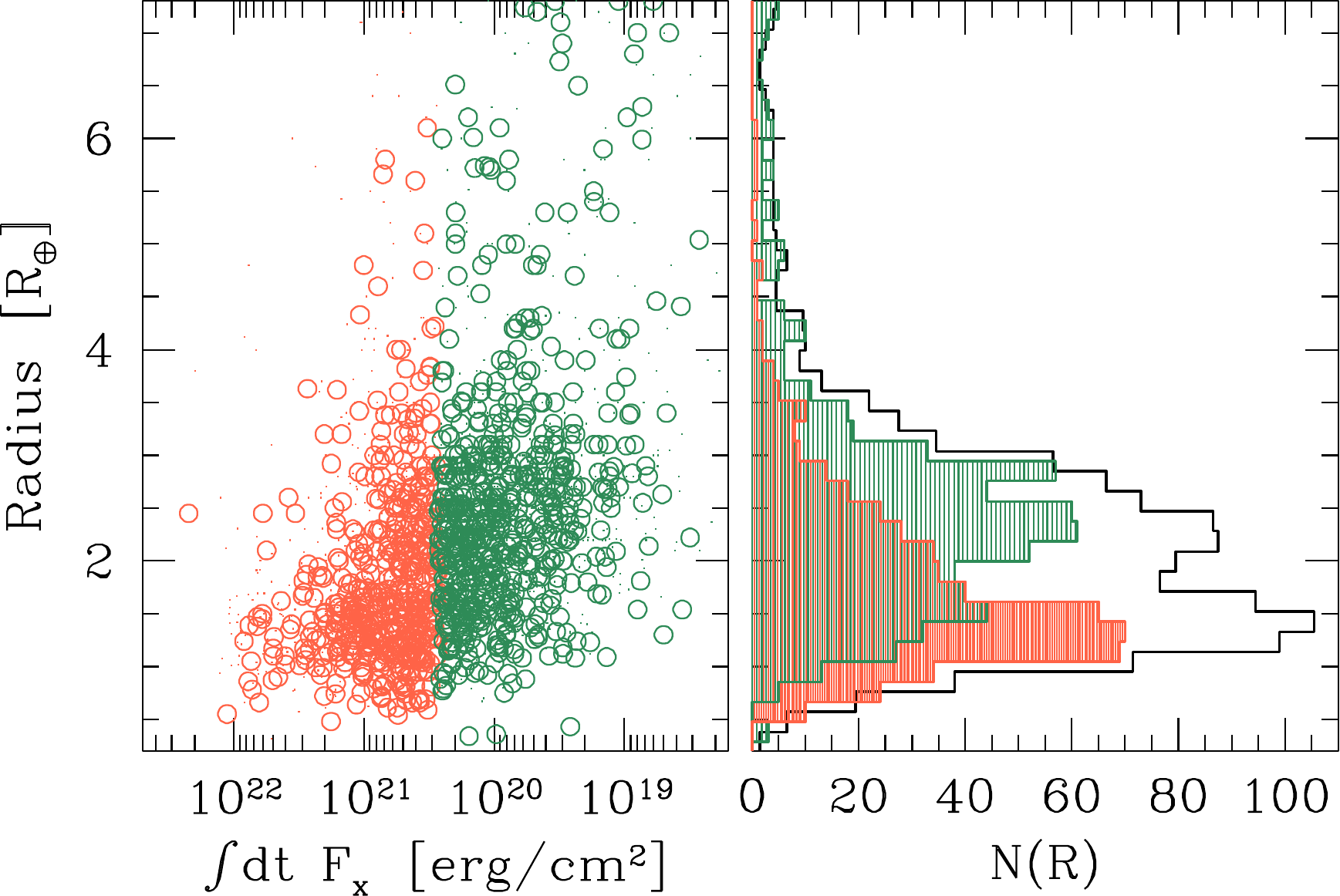}
\caption{The correlation between radii of KOI and their X-ray
  exposures.  In the left panel, we show the radii of KOIs in
  multi-planet systems (in open circles), {\bc and in KOI singles
    (dots)}, versus their integrated X-ray exposure.  The division
  between the red and the green population corresponds to a distance
  of $0.1$ AU around a sun-like star. The black solid histogram in the
  right panel is the size distribution of all KOI multi-planet systems
  {\bc (dashed histogram for singles)}, while green that of the low
  X-ray group and red the high X-ray group.  The red and green objects
  have distinctly different peak sizes, as is expected from our X-ray
  evaporation models (Figure~\ref{fig:rad_sep}). While the green
  objects (with a peak size at $\sim 2.5 R_\oplus$) manage to hold on
  to some of the hydrogen envelopes they were born with, most of the
  red objects have been stripped down to naked cores (with a peak size
  at $ \sim 1.5 R_\oplus$). The bimodal size distribution of KOIs is
  naturally explained by their mass-loss history (see text).}
\label{fig:wu2}
\end{figure}

In Figure \ref{fig:wu2}, we demonstrate that the radius distribution
of KOI multi-planet systems {is suggestive of a bi-modal distribution}. Most planets have sizes $\sim
1.5 R_\oplus$ or $2.5 R_\oplus$, with a deficit of objects at radius
$\sim 2 R_\oplus$, {\bc indicative of the presence of a gap}. To {\bc further test} this
suspicion, we divide the objects by their X-ray exposure into a high
X-ray group (corresponding to $< 0.1$ AU around a solar-type star) and
a low X-ray group. Strikingly, objects with a high X-ray exposure
mostly have sizes below the gap, at $\sim 1.5 R_\oplus$, while objects
with a low X-ray exposure typically have sizes above the gap, at $\sim
2.5 R_\oplus$. This argues that the deficit at $2 R_\oplus$ {could be}
physical and is associated with X-ray exposure.

The same bi-modal behaviour
appears when we consider only KOI singles, {\bc both single and multi-planet} KOIs, or
when we include only bright KOI targets, or only dim KOI targets. {\y 
  In addition, the same behaviour is seen when we restrict ourselves to
  planets that have periods shorter than $50$ days and sizes above
  $1.3 R_\oplus$, a group of KOIs that suffer relatively little
  incompleteness effect \citep{Petigura,Fressin}.} {\y 
  Although most careful studies to date have yet to have sufficient
  radius resolution to confirm this feature
  \citep{Howard,Petigura,Fressin},  it is hinted at by \citet{morton} who construct a probability distribution function rather than using histograms with large ranges.}
{\y Currently evaporation is the only process that can naturally explain
  such a bimodality. Any other processes (planet gas accretion,
  migration, orbital instability, planetary mergers) may lead to a
  correlation between planet size and location, but will not produce
  two separate peaks in radius.} 
  
  The presence of this gap provides strong evidence that evaporation not
only sculpts the upper envelope of planet sizes, it is also driving
the evolution of the majority of KOI objects. {\bc If all the planets in the higher X-ray exposure peak originally had significant H/He envelopes comparable to the planets with lower X-ray exposures, then $\sim$ 50\% of
{\it Kepler} planet candidates having experienced significant mass-loss
during their lifetimes}.  Comparing the gap location ($0.1$ AU around
sun-like star) against our theoretical calculations, we suggest that
the planet population {\y in the current KOI list}\footnote{ \y The
  apparent absence of bare rocky planets at large separations deserves
  a comment. The current KOI list is incomplete for this population
  \citep{Howard,Fressin,Petigura}. As such, their presence is
  not yet understood. } have predominately low mass cores $\sim 6$
M$_\oplus$, and that most started out their lives with Hydrogen/Helium
envelopes of at least a few percent in mass. {\bc Certainly, the radius distribution of close-in planets requires further work - along the lines of Morton et al. (2013) - before definitive conclusions can be drawn. As an accurately determined bi-modal distribution encodes value information about the initial and final planet mass and composition.}
%
%
Figure~\ref{fig:rad_sep} illustrates that the gap radius, as well as
the separation at which this gap appears, are direct probes of the
core properties. An improved investigation on core composition and
mass should be conducted when planet radii are better determined.

\subsection{Comparison of Planet Density}
\label{subsec:density}

\begin{figure}
\centering
\includegraphics[width=\columnwidth]{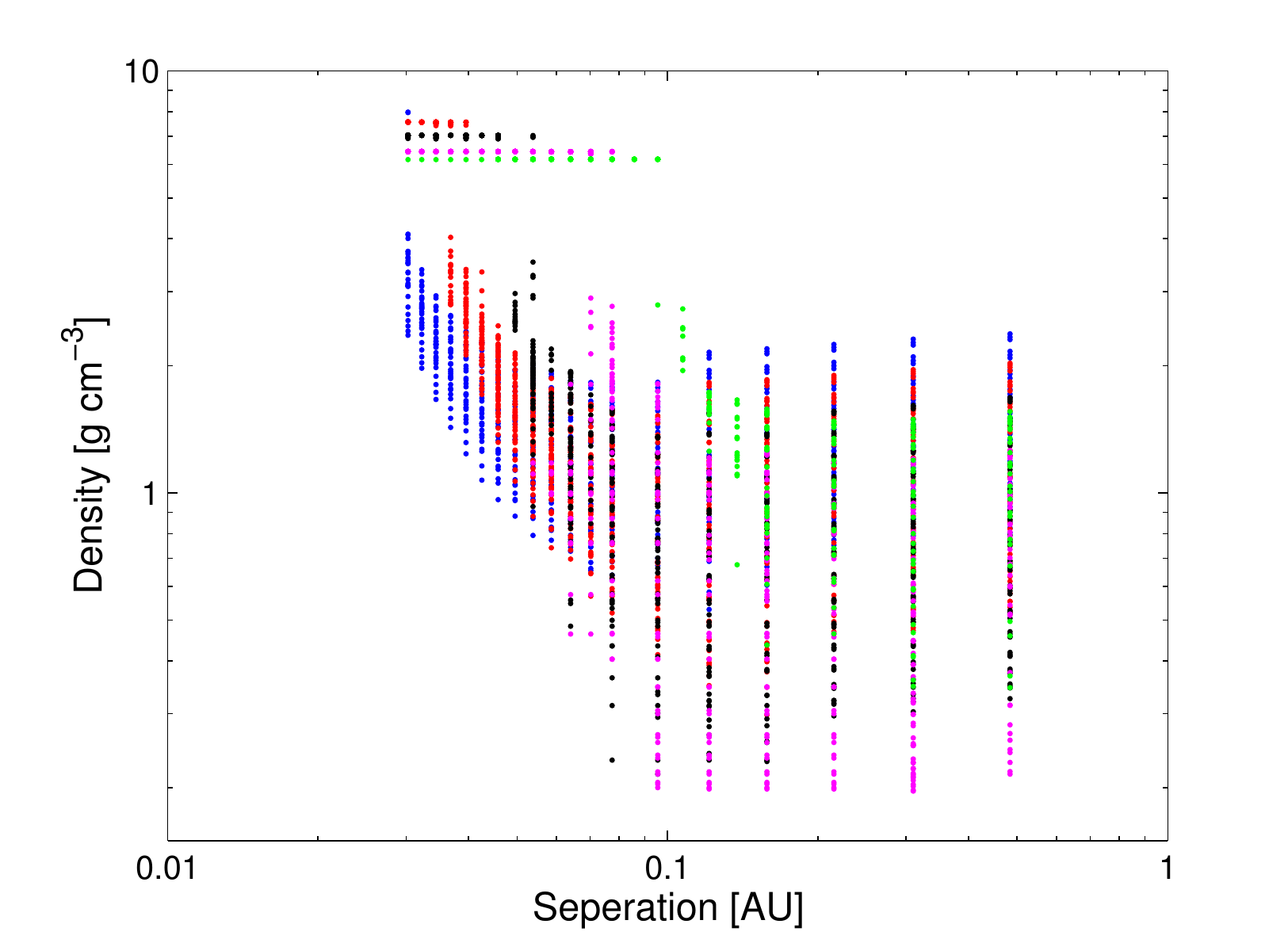}
\caption{Same as Figure~\ref{fig:rad_sep} but showing final planet density.}
\label{fig:den_sep}
\end{figure}

\begin{figure}
\centering
\includegraphics[width=\columnwidth]{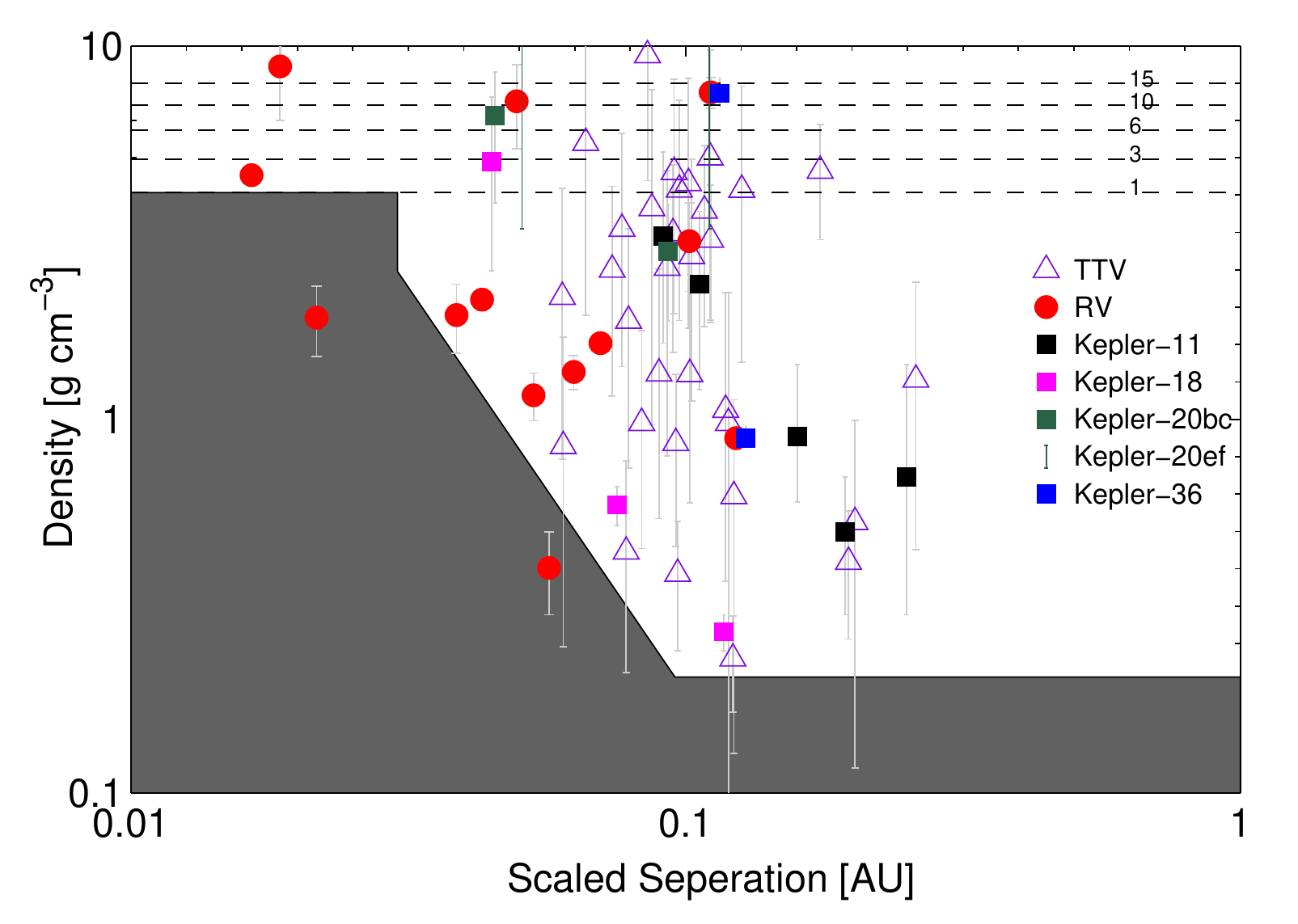}
\caption{Observed planet density plotted against orbital separation,
  scaled to have the same X-ray exposure as around a sun-type
  star. The planet sample includes both those from radial velocity
  surveys \citep[red filled circles, as compiled by][]{Weiss} and
  Kepler planets with masses confirmed by the {\it Kepler} team
  \citep[compiled from][]{Lissaueretal11,kepler18,kepler20,kepler36}
  or calculated from transit timing variations (TTV) by
  \citet{WuLithwick}.  There are large error bars to many of the TTV
  masses.  The shaded region is the disallowed region as shown in
  Figure~\ref{fig:den_sep}, and the dashed lines show the densities of
  bare rocky cores for a range of masses.}\label{fig:den_compare}
\end{figure}

While the KOI catalogue only allows comparison of planet radius, the
small but growing sample of low-mass planets with measured masses also
provides another important comparison: planet density.  In
Figure \ref{fig:den_sep}, we illustrate planet densities resulting from
our integration. In the density-separation plane, the upper envelope
in planet size is now translated into a lower envelope in planet
density. There is a gap in planet density, similar to that in radius.
However, if there is a planet population with low mass cores ($1-3
M_\oplus$), these low density cores may partially fill in the gap.

To compare, in Figure \ref{fig:den_compare} we plot the measured
densities of a list of low masses planets against their scaled
separations. We scale the actual planet separation by their respective
X-ray exposure, though since most host stars are solar-type, this
correction is typically small ($<10\%$).  The theoretically disallowed
low-density region is shown. Most of the observed densities avoid this
forbidden region, providing strong evidence for sculpting by
evaporation.

Current density measurements (especially those using transit-timing
fittings) contain large uncertainties. This prevents us from making
more quantitative comparisons at the moment. In particular, we could
not discern the density gap as predicted by model calculations.

The particular case of Kepler-36bc \citep{kepler36} is worth
commenting.  The two planets have nearly identical separations (0.115
AU and 0.128 AU) but largely discrepant densities, 6.8 \& 0.86 g
cm$^{-3}$, respectively.  At their present orbits, the minimum core
masses for the two planets to retain their envelopes is $\sim 6.5$
M$_\oplus$ (see Figure \ref{fig:den_sep}).  The measured masses of the
two planets are 4.45 $M_\oplus$ and 8.08 $M_\oplus$, respectively,
naturally explaining the diversity in their structure.  More systems
like Kepler-36 will be able to provide strong constraints on the
nature and strength of evaporation in close-in planetary systems.


\section{Discussions}\label{sec:caveats}

  We discuss some of the caveats
and limitations of the presented calculations and  how they bear
 on our inferences   about the observed planet
population.   The two most important assumptions
  concern the use of a two-layer planetary model (a rock
core plus a Hydrogen/Helium envelope), and the  
  adoption of the \citet{owen} evaporation model.

\subsection{Variations in the planetary structure}

  Low-mass planets could contain volatile-rich atmospheres, they could
  also contain a significant amount of iron and/or ice in their cores,
  \citep{degeneracy,rogers_seager}.  Our simple two-layer model of
  rocky cores plus hydrogen envelope has been successful in
  explaining the observations. But what about these other
  possibilities? Can we exclude them based on current observations?

  We can exclude the possibility that the dominant primordial atmospheres of
  these low-mass planets are very rich in volatiles. Our arguments
  below run similarly to those given in \citet{WuLithwick} but are
  more informed by our detailed modelling and by the physics of
  evaporation. If one considers a water-rich envelope, for
  instance, evaporation will not proceed as is described here. First,
  water molecules have to be photo-dissociated, then oxygen has to
  settle out to produce a nearly pure hydrogen upper
  atmosphere \citep[e.g.][]{kasting1983}. If oxygen is present in the
  evaporative flow at a high enough concentration, it will produce
  strong cooling and increase the opacity to the X-rays. This will severely suppress the gas temperature, leading to
  a much lower evaporation rate, similar to what occurs in a high
  metalicity protoplanetary discs \citep{EC10}, but more extreme. Now suppose all these
  conditions are satisfied and all hydrogen in the water atmosphere is
  lost, this removes $1/8$ of the atmosphere mass. However, since the
  original water atmosphere has a low scale height, such a removal can
  hardly change either the bulk size or the bulk density of the
  planet. One would therefore not be able to explain the upper
  envelope in the observed planet radius, or the bimodal planet size
  distribution, or the correlation between planet density and X-ray
  exposure in terms of planetary evaporation.


  In contrast to atmosphere composition, we can be less certain about
  the core composition. For planets with hydrogen envelopes that are
  more than a percent in mass, the planet sizes are not sensitive to
  the core sizes (or equivalently, core composition). Planets that
  have been evaporated down to bare cores may be able to inform us on
  the core composition, if the core masses are known. A number of
  these objects show densities that are compatible with rocky or
  iron/rock compositions \citep[e.g.,][]{corot7b,kepler10b}. More
  detailed investigations are necessary to ascertain the core
  compositions.


\subsection{Improving the evaporation model}

As we have discussed previously, evaporation is key to the evolution of close in planets. Thus it is important to assess the role the assumed evaporation model plays in our conclusions.
\begin{figure}
\centering
\includegraphics[width=\columnwidth]{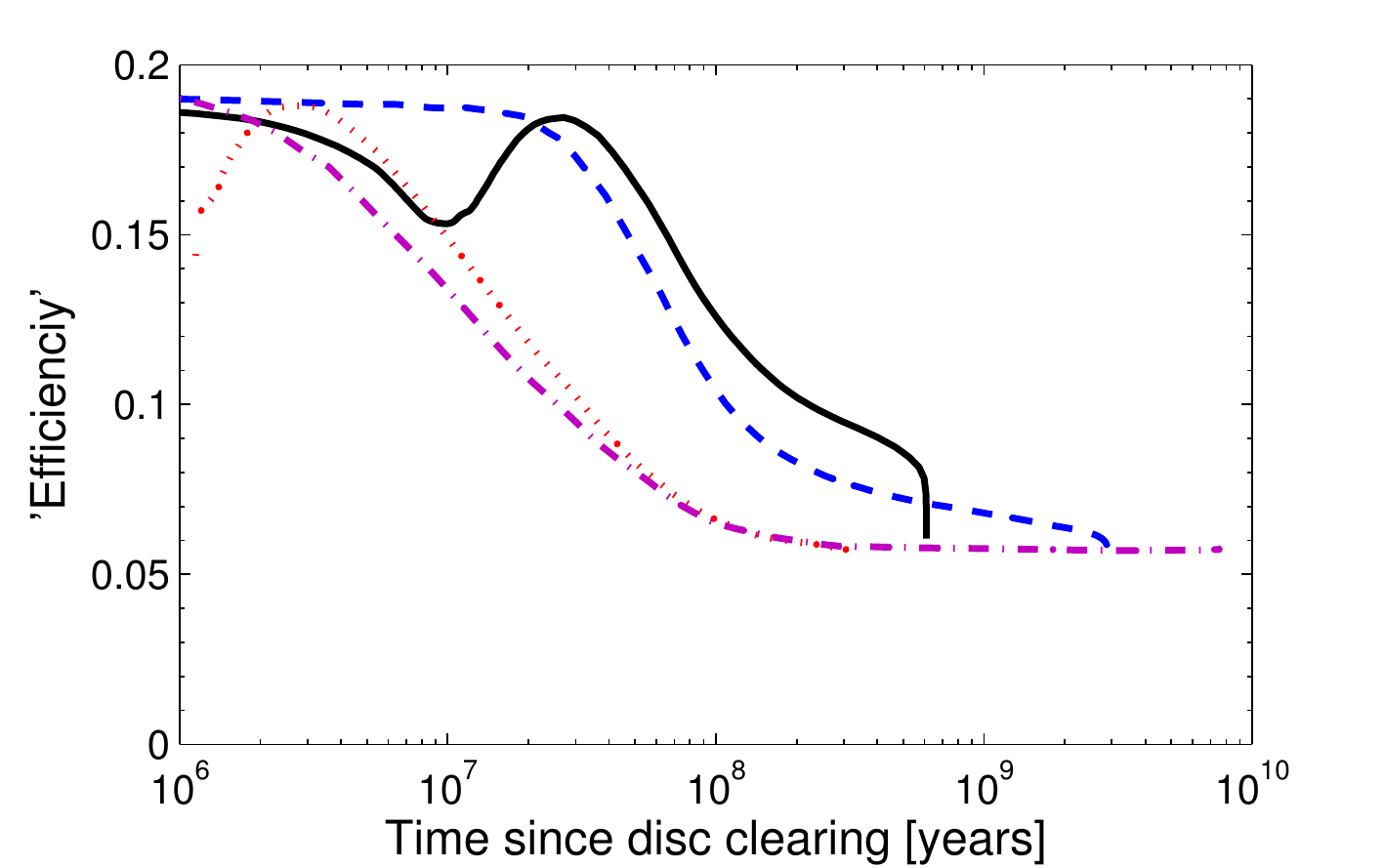}
\caption{The evaporation `efficiency' ($\eta$) as a function of
  time. Here, we trace the evolution of four $20$ M$_\oplus$ planets,
  with core masses of 7.5 (solid), 10 (dashed), 12.5 (dotted) \& 15
  (dot-dashed) M$_\oplus$, at a separation of $\sim$0.03 AU from a sun-like
  star. All models lose their
  entire envelopes towards the end.}
\label{fig:eff}
\end{figure} Most previous attempts to model the evolution of evaporating planets
use a constant evaporation efficiency ($\eta$), which is typically  taken to be $\sim 10\%$
\citep[e.g.,][]{lopez}. We have argued that this efficiency depends on
planet mass and radius, as well as on the X-ray flux. We further
demonstrate this point here by showing how the efficiency changes as a
planet evolves in Figure~\ref{fig:eff}.
Following four planets with the same initial mass of $20 M_\oplus$
(but different core masses), we find that $\eta$ can decrease by a
factor of $4$ as the planets evolve from the early puffy stage to the
later denser stage, though the variations are not strictly monotonic
in time. The overall decrease can be understood: as the planet evolves
due to mass-loss and thermal contraction, the planet's density and the
surface escape velocities increases with time. Therefore, it takes
longer to accelerate the flow to the escape velocity. This leads to
enhanced cooling and lower efficiencies. {\y The non-uniform evolution of some of the models at early times is due to planets straddling the peak efficiency line (roughly when $T_{\rm gas}\sim T_{\rm escape}$, see Owen \& Jackson 2012) during their evolution and moving above and below it at early times. }


While the fixed efficiency of $\sim 10\%$ adopted by, e.g.,
\citet{jackson2012,lopez} does represent the approximate median value
during the evolution, it can result in order unity inaccuracies in the
integrated mass-loss. So any inferences about the initial planet
structure should be taken with caution.

\subsubsection{Accuracy of the X-ray model}

The \citet{owen} model for X-ray evaporation contain a number of
assumptions that may impact our conclusions here.
First, since only the soft $< 1-2$ keV photons are responsible for
heating, while the X-ray flux refers to the entire observed X-ray spectrum ($0.1-10$ KeV).
If the adopted X-ray spectrum -while based on observed spectra
\citep[see][]{owen10}- is overly soft or overly hard, then the X-ray
heating will be over- or under-estimated, respectively.
%
Second, these calculations assumed that, in the X-ray region,
molecules are photo-dissociated or thermally dissociated by UV
photons. Therefore, Owen \& Jackson (2012) neglected cooling associated with
molecular species. This is an important assumption which has not been
vigorously tested.  If instead the molecular species provide
significant cooling (in the temperature range of $ 2000-5000$K), the
X-ray driven flow will not reach $T > 5000$K and may remain sub-sonic
all the way out to the EUV ionization-front. This will markedly reduce
the mass-loss rate. More detailed calculations are necessary to
address this issue in the future.

\subsubsection{The role of EUV Evaporation}

\begin{figure}
\centering
\includegraphics[width=\columnwidth]{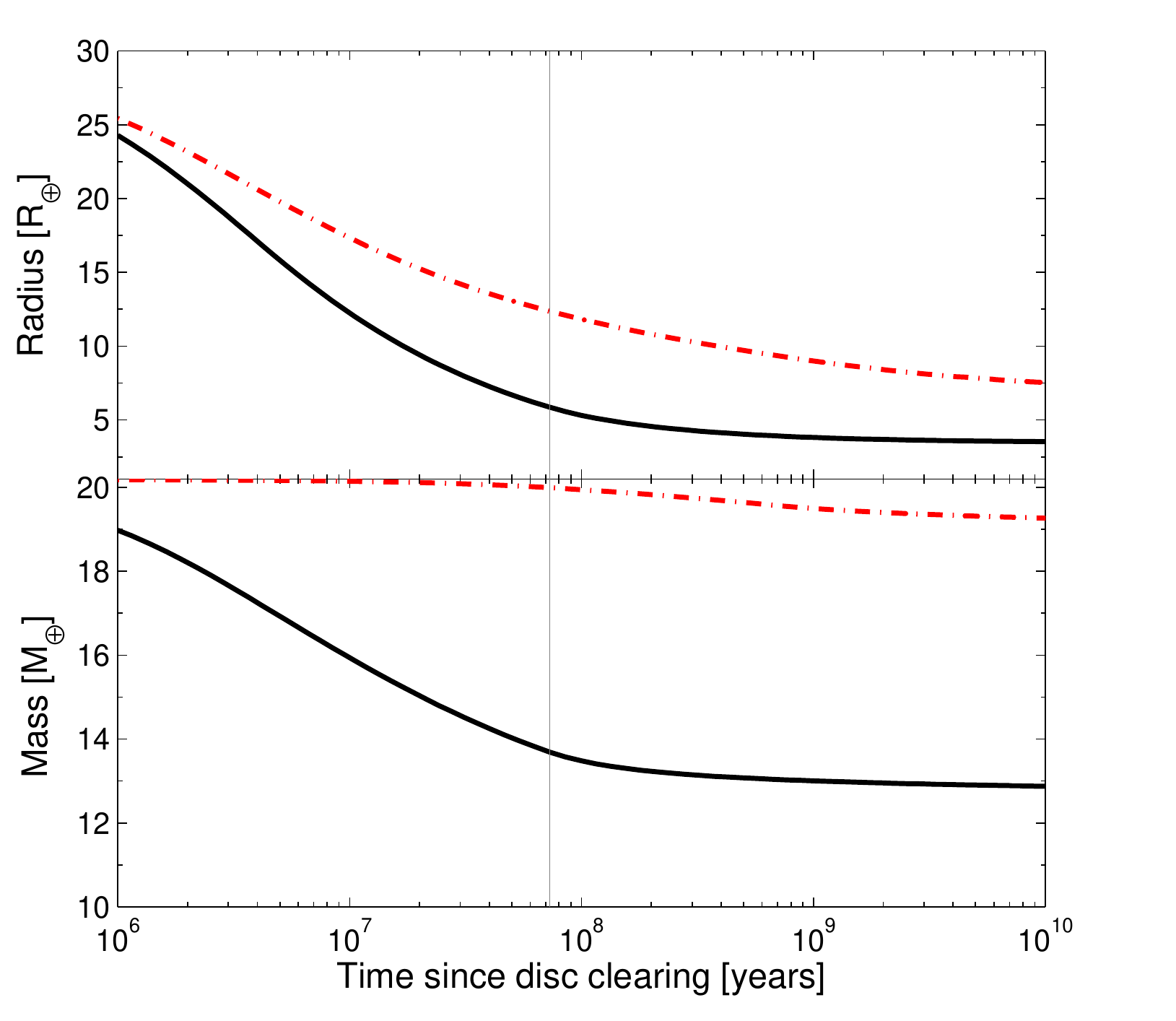}
\caption{The mass and radius evolution of a 20 M$_\oplus$ planet with
  a 12.5 M$_\oplus$ core at a separation of $\sim 0.05$ AU undergoing standard Owen \& Jackson (2012)
  evaporation (solid) -same as shown in Figure~\ref{fig:standard}- 
  and pure EUV evaporation (dashed). The thin solid line represent the
  time at which the X-rays/EUV begin to decline.
}\label{fig:evolve_euv}
\end{figure}

We find that EUV evaporation contributes $<10\%$ of the total
mass-loss during the planet's evolution. The EUV portion of the
stellar flux, though comparable in energy to that of the X-rays, are
much less efficient at driving a high {\it integrated} mass-loss. This
is because in the EUV flow region, the recombination of hydrogen and
the subsequent cooling bleeds much of the energy to space. The
radiation hydrodynamics is similar to that of a HII region where
ionization is balanced by recombination and the temperature is
controlled by the cooling thermostat to $\sim 10^4$ K
\citep{murrayclay2009}.


We re-calculate the evolution of our `standard' model 
(see Section~\ref{sec:standard})
undergoing EUV evaporation only, the results of which are shown in Figure~\ref{fig:evolve_euv}.
At high EUV fluxes ($\ga10^4$ erg s$^{-1}$), the mass-loss scales as
$L_{\rm EUV}^{1/2}/a$ \citep{murrayclay2009,owen} and is significantly lower than
the  values that apply to X-ray flow.
This suppresses the mass-loss at early times by a factor of $\sim 10$,
relative to the X-ray model. The integrated mass-loss is $\sim 1$
M$_\oplus$, as opposed to $\sim 7$ M$_\oplus$ in the X-ray model.  As
such, EUV evaporation alone is unable to sculpt the close-in planet
population and cannot explain the observed features discussed in
Section~\ref{sec:pop}.


\section{Conclusions}

In this work we have coupled the hydrodynamic evaporation models of
\citet{owen} to the stellar evolution code {\sc mesa}, in order to
follow the mass and radius evolution of low mass planets orbiting
close to their stars.  Evaporation, while having little effect on
massive planets, can remove the entire Hydrogen envelopes for the
hottest low-mass planets.  In all cases we find that X-rays is the
dominant sculpting force, and that most of the mass-loss occurs in the
first few $100$ Myrs when the stars were more chromospherically active
and while the planets were still contracting thermally.

Our main conclusions are as follows:

\begin{enumerate}
\item Evaporation produces an upper envelope in planet radius as a
  function of separation. The location of this upper envelope depends
  on planet masses and the X-ray luminosity of the host stars.  In
  particular, M-dwarfs, having proportionally larger X-ray fluxes,
  should have stronger evaporative power than indicated by their low
  bolometric luminosities.

\item To closely reproduce the observed upper envelope in {\it Kepler}
  candidates, both around sun-like stars and around stars of other
  spectral types, we require that the most massive hot Neptunes should
  have total masses not much exceeding $\sim 20$ M$_\oplus$, and core
  masses roughly half of that.

\item Very close-in planets can be stripped of their entire
  atmosphere. The boundary between complete loss and planets that can
  retain at least $\sim 1\%$ of their atmosphere lies at $\sim 0.1$ AU
  if the core masses are $\sim 6 M_\oplus$. At this distance (where
  most of the {\it Kepler} planets lie), a thinner envelope can not
  survive. So we expect a gap in planet size distribution -- and this
  is {\bc suggestively seen} in the Kepler catalogue, where there is a deficit of
  planets at $\sim 2 R_\oplus$ straddling planets with high X-ray
  exposures and those with low X-ray exposures.
  Comparison with our models suggests that most of the {\it Kepler}
  planets should have core masses $\sim 6 M_\oplus$ and should have
  primordial Hydrogen/Helium envelopes at least a percent in
  mass. Moreover, about half of the {\it Kepler} planets belong to the
  high X-ray group and have been stripped down to naked cores.

\item Evaporation naturally explains the observed correlation between
  planet density and separation. At closer separations, planets in
  general have higher densities. 

\item Combining all evidences that support evaporaton, we argue that
  Kepler planets were born with hydrogen envelopes, not volatile-rich
  atmospheres \citep[also see][]{WuLithwick}.

\end{enumerate}

Looking ahead, we expect that the approach we adopt here, coupling
thermal evolution and evaporation, will perhaps be the only hope we
have for recovering the initial structure of low-mass planets. With
better determined stellar radii and hence more accurate planet radii,
we may be able to retrieve the initial distribution of planet total
masses and core masses. With more measurements of planet densities, it
may be possible to reconstruct the histories of individual planets
\citep[as is done for Kepler-11 by][]{lopez}.  A larger sample of
planets with measured masses and radii will allow us to place
constraints on the core compositions, as well as initial planet
entropy.
It is at this point that we will begin to learn valuable information
about the planet population at birth and make inferences about the
planet formation process
\citep[e.g][]{ida2005,mordasini2012b,mordasini2012a,hansen2012,hansen2013}.\\

\acknowledgments

We are grateful to the anonymous referee for comments that helped
improve the paper. We thank Norman Murray, Chris Thompson, Jason Rowe,
Alan Jackson, Eric Lopez, Adam Burrows and Yoram Lithwick for
interesting discussions. The calculations were performed on the
Sunnyvale cluster at CITA which is funded by the Canada Foundation for
Innovation. YW acknowledges support by NSERC and the Province of
Ontario.

\bibliographystyle{apj}

\bibliography{orig}

\end{document}